\pdfoutput=1
\documentclass[twocolumn,prl,superscriptaddress,longbibliography]{revtex4-2}
\usepackage{graphicx,amsmath,amssymb,tikz,scalerel}
\usepackage[colorlinks=true, citecolor=blue, urlcolor=blue, linkcolor=red]{hyperref}
\renewcommand{\section}[1]{{\par\it #1---}\ignorespaces}
\usetikzlibrary{svg.path}
\definecolor{orcidlogocol}{HTML}{A6CE39}
\tikzset{orcidlogo/.pic={
		\fill[orcidlogocol] svg{M256,128c0,70.7-57.3,128-128,128C57.3,256,0,198.7,0,128C0,57.3,57.3,0,128,0C198.7,0,256,57.3,256,128z};
		\fill[white] svg{M86.3,186.2H70.9V79.1h15.4v48.4V186.2z}
		svg{M108.9,79.1h41.6c39.6,0,57,28.3,57,53.6c0,27.5-21.5,53.6-56.8,53.6h-41.8V79.1z M124.3,172.4h24.5c34.9,0,42.9-26.5,42.9-39.7c0-21.5-13.7-39.7-43.7-39.7h-23.7V172.4z}
		svg{M88.7,56.8c0,5.5-4.5,10.1-10.1,10.1c-5.6,0-10.1-4.6-10.1-10.1c0-5.6,4.5-10.1,10.1-10.1C84.2,46.7,88.7,51.3,88.7,56.8z};}}
\newcommand\orcid[1]{\href{https://orcid.org/#1}{\mbox{\scalerel*{\begin{tikzpicture}[yscale=-1,transform shape]\pic{orcidlogo};\end{tikzpicture}}{|}}}}

\begin{document}
\title{Restoring Quantum Superiority of Noisy Quantum Illumination}
\author{Wei Wu\orcid{0000-0002-5989-9458}}
\affiliation{Key Laboratory of Quantum Theory and Applications of Ministry of Education, Lanzhou Center for Theoretical Physics, Gansu Provincial Research Center for Basic Disciplines of Quantum Physics, and Key Laboratory of Theoretical Physics of Gansu Province, Lanzhou University, Lanzhou 730000, China}
\author{Jun-Hong An\orcid{0000-0002-3475-0729}}
\email{anjhong@lzu.edu.cn}
\affiliation{Key Laboratory of Quantum Theory and Applications of Ministry of Education, Lanzhou Center for Theoretical Physics, Gansu Provincial Research Center for Basic Disciplines of Quantum Physics, and Key Laboratory of Theoretical Physics of Gansu Province, Lanzhou University, Lanzhou 730000, China}

\begin{abstract}
Quantum illumination uses quantum entanglement as a resource to enable higher-resolution detection of low-reflectivity targets than is possible with classical techniques. This revolutionary technology could transform modern radar. However, it is widely believed that the decoherence induced by the ubiquitous quantum noise destroys the superiority of quantum illumination, severely constraining its performance and application in our present noisy intermediate-scale quantum era. Here, we propose a method to restore the quantum superiority of the quantum illumination in the presence of quantum noises. Going beyond the widely used Born-Markov approximation, we discover that the resolution of noisy quantum illumination is highly sensitive to the energy spectrum of the composite system formed by each of the two light modes and its local quantum noise. When a bound state is present in the energy spectrum, the resolution asymptotically approaches its ideal form. Our result establishes a physical principle to preserve the quantum superiority and paves the way for the realization of high-resolution quantum illumination in noisy situations.
\end{abstract}
\maketitle

\section{Introduction}\label{sec:sec1}
A second quantum revolution is underway \cite{doi:10.1098/rsta.2003.1227}. It uses quantum resources to develop revolutionary technologies with unprecedented levels of performance. As quantum technologies become more widely used, people are becoming more adept at detecting, processing, and securing information. Quantum illumination is a notable example of this emerging technology. It uses quantum entanglement to outperform the corresponding classical benchmarks for target detection~\cite{doi:10.1126/science.1160627,PhysRevLett.101.253601,Shapiro_2009}. Its main idea is to generate two entangled light beams, called the signal and the idler, respectively. The signal beam is sent to detect possible targets in the region of interest, while the idler beam is retained at the source, awaiting recombination with the signal upon its eventual return \cite{Blakey2022,Liu2023,PhysRevX.14.041058,10.1063/5.0162419,doi:10.1126/sciadv.aay2652,PhysRevLett.118.070803,PhysRevLett.127.173603,PhysRevLett.127.040504,PhysRevLett.131.033603,Volkoff_2024,PhysRevLett.118.040801,PhysRevApplied.20.024030,PhysRevA.107.062405,PhysRevLett.133.130801}. A joint measurement of the pair is performed to capture information held by their entangled nature, which enables an improved resolution in target detection compared to classical illumination using coherent beams \cite{PhysRevLett.119.120501,doi:10.1126/science.1160627,PhysRevLett.101.253601,Shapiro_2009,PhysRevA.108.062605,PhysRevA.109.062440}. Recently, quantum illumination has been experimentally realized by using the Josephson parametric converter at microwave frequencies~\cite{doi:10.1126/sciadv.abb0451,10.1063/1.5085002,8890864,Assouly2023}. This provides a building block for developing transformative radar techniques \cite{Zhuang2023ER,GALLEGOTORROME2024100497,Karsa_2024}.  

Quantum illumination is on the way to enhancing its practical capacity to outperform state-of-the-art classical counterparts. However, we are still in the noisy intermediate-scale quantum era, where ubiquitous quantum noise in the microscopic world is not under good controllability. The decoherence of quantized lights caused by various types of quantum noise is the primary source of errors in many optical or microwave tasks~\cite{PhysRevLett.128.010501,Reichert2022,PhysRevLett.133.130801,Karsa_2024,10872958}. Previous studies~\cite{PhysRevA.90.052308,PhysRevLett.114.110506,Jonsson_2022,Park2023,PhysRevResearch.5.033010,PhysRevA.111.022627} have shown that the detection error of quantum illumination increases rapidly once the decoherence of the involved quantized lights is considered. This result reveals that the superiority of quantum illumination enabled by quantum entanglement is generally destroyed by decoherence. Thus, achieving the promised quantum advantage in the realistically noisy situation is one of the major challenges in realizing a high-resolution quantum illumination. However, a clear imperfection leading to the above result is that it is based on the Born-Markov approximation to describe the decoherence. Thus, the determination of whether the destruction effect of the decoherence on quantum illumination is ostensible or fundamental and whether it can be overcome or not is highly desirable from both theoretical and experimental perspectives.

\begin{figure}
\centering
\includegraphics[height=0.35\columnwidth]{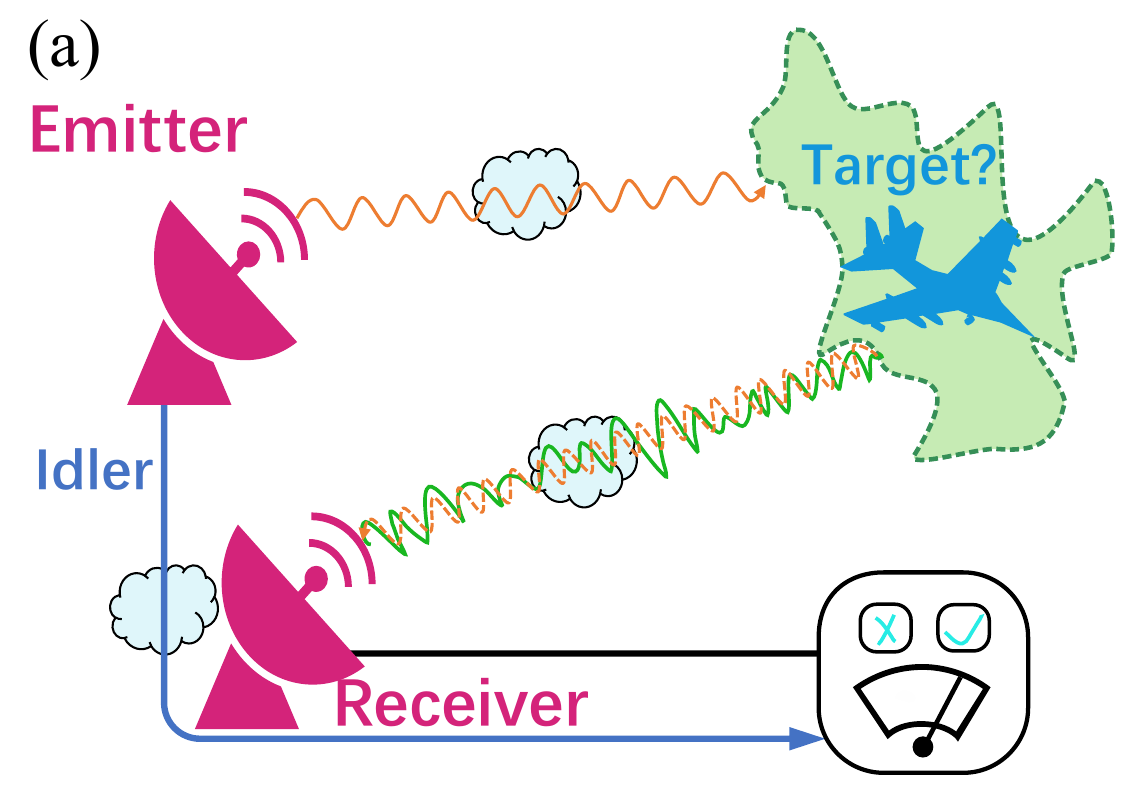}
\includegraphics[height=0.35\columnwidth]{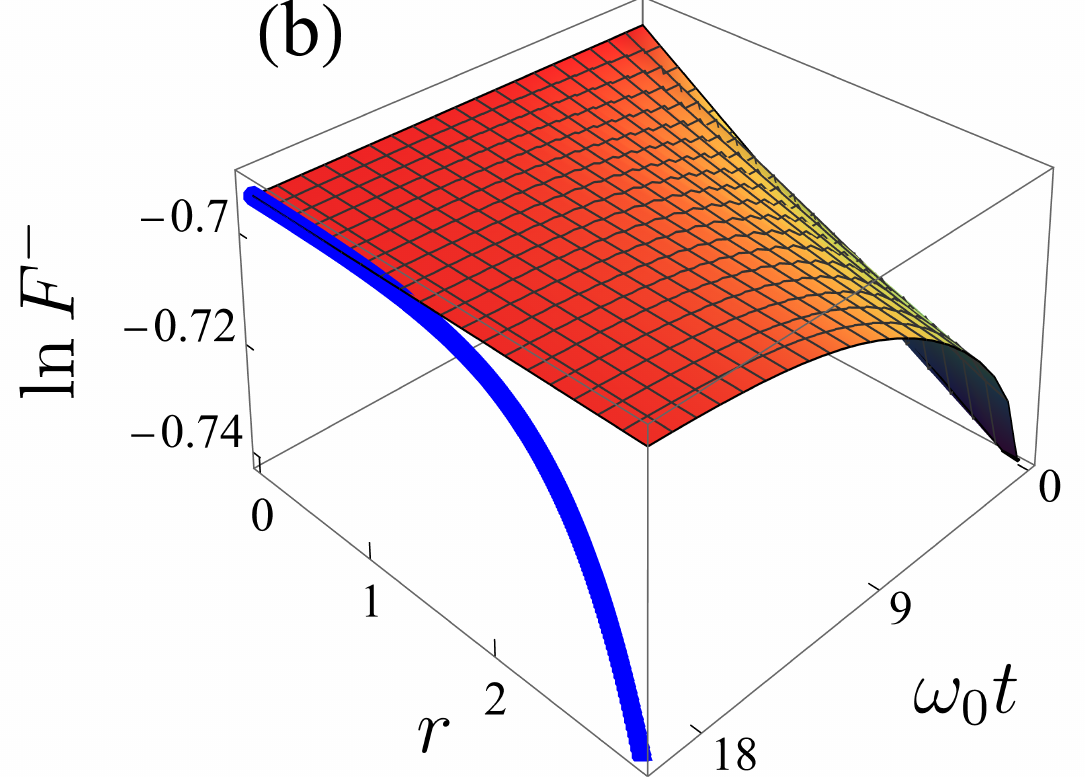}
\caption{(a) Diagrammatic sketch of the quantum illumination scheme. (b) Evolution of the resolution characterized by the lower bound $F^-(t)$ of the minimum error probability $P(t)$ in different values of squeezing parameter $r$ under the Born-Markovian decoherence. The blue line is the ideal result. We use $\omega_{c}=10\omega_{0}$, $\eta=0.05$, $\xi=10^{-3}$, $s=0.8$, and $\beta=\omega_{0}^{-1}$.}\label{fig:fig1}
\end{figure}

Here, we propose a noise suppression scheme to protect the quantum superiority of quantum illumination from the destruction of individual decoherence of the two quantized light beams. Going beyond the Born-Markov approximation to describe the decoherence, we discover that the resolution of the quantum illumination sensitively depends on the feature of the energy spectrum of the composite system consisting of each of the two light beams and its local noise. Accompanying the formation of a bound state in the energy spectrum, the resolution asymptotically returns to its ideal behavior. Overcoming the decoherence problem faced by quantum illumination, our result provides a useful guidance to beat the noisy effects on quantum illumination and pave the way for the development of quantum radar in the current noisy intermediate-scale quantum era.

\section{Ideal scheme}\label{sec:sec2}
We first consider an ideal scheme of quantum illumination without the decoherence induced by quantum noise. The quantum entanglement is used as a resource to boost the resolution of target detection [see Fig.~\ref{fig:fig1}(a)]. In the first step, two quantized light beams are prepared in an entangled two-mode squeezed vacuum state $|\psi(0)\rangle=\exp[r(\hat{a}_{\text{A}}\hat{a}_{\text{B}}-\hat{a}^{\dagger}_{\text{A}}\hat{a}^{\dagger}_{\text{B}})]|00\rangle$, where $r$ is the squeezing parameter and $\hat{a}_\text{A/B}$ are the annihilation operators of the two light beams, by the technique of spontaneous parametric down-conversion at the optical frequencies~\cite{PhysRevA.57.3123,PhysRevLett.99.243601,PhysRevA.92.023804,PhysRevX.9.031033} or Josephson parametric converter at the microwave frequencies~\cite{doi:10.1126/sciadv.abb0451,10.1063/1.5085002,8890864,Assouly2023}. In the second step, the mode-A light beam acting as the signal is irradiated to the region in which a low-reflectivity target might exist. The mode-B light beam acting as the idler is retained at the source for later joint measurements with the reflected signal mode to judge whether the target is present or absent. If the target is absent, then the mode-A photons would not be reflected, and only the thermal photons from the background are captured by the detector. Combined with the idler mode B, the state received by the detector is
\begin{equation}
\rho_{0}(t)={\exp(-\beta \omega_{\text{b}}\hat{a}^{\dagger}_{\text{b}}\hat{a}_{\text{b}})\over\text{Tr}[\exp(-\beta \omega_{\text{b}}\hat{a}^{\dagger}_{\text{b}}\hat{a}_{\text{b}})]}\otimes \text{Tr}_{\text{A}}[\rho(t)].
\end{equation}
Here, $\beta=T^{-1}$ is the inverse temperature of the background photons with annihilation operator $\hat{a}_\text{b}$ and frequency $\omega_\text{b}$ and $\rho(t)=\exp(-i\hat{H}_{\text{s}}t)|\psi(0)\rangle\langle \psi(0)|\exp(i\hat{H}_{\text{s}}t)$ is the evolved state of the two light beams governed by their free Hamiltonian $\hat{H}_{\text{s}}=\sum_{l=\text{A},\text{B}}\omega_0\hat{a}_{l}^{\dagger}\hat{a}_{l}$. We have used $\hbar=k_\text{B}=1$. Without loss of generality, we assume $\omega_{\text{b}}=\omega_{0}$. In contrast, if the target is present, then the mode-A photons have a probability to be reflected by the target. Thus the state received by the detector becomes
\begin{equation}
\rho_{1}(t)=\xi \rho(t)+(1-\xi)\rho_0(t),
\end{equation}
where $\xi$ denotes the reflectivity of the target and generally satisfies $0<\xi\ll 1$. The resolution of the quantum illumination is characterized by the discrimination between $\rho_{0}(t)$ and $\rho_{1}(t)$ defined as the minimum error probability
$P=\big{[}1-2^{-1}\text{Tr}|\rho_{1}(t)-\rho_{0}(t)|\big{]}/2$, where $\text{Tr}|\rho_{1}(t)-\rho_{0}(t)|$ is the trace norm of $\rho_{1}(t)-\rho_{0}(t)$ measuring their distance \cite{PhysRevA.78.012331,RevModPhys.84.621,PhysRevA.110.052403}. This is called the Helstrom limit. However, it is generally difficult to obtain the analytical expression of $P$~\cite{PhysRevLett.133.110801}. Fortunately, its lower bound can be constructed by the quantum fidelity $F(\rho_{0},\rho_{1})\equiv[\text{Tr}(\sqrt{\rho_{0}}\rho_{1}\sqrt{\rho_{0}})^{\frac{1}{2}}]^{2}$ between $\rho_0(t)$ and $\rho_1(t)$ as $F^-\equiv\frac{1}{2}[1-\sqrt{1-F(\rho_{0},\rho_{1})}]$~\cite{PhysRevA.77.032311,PhysRevLett.115.260501}. Sharing the same monotonicity with $P$, the lower bound $F^-$ itself can be used as a measure of the performance of the quantum illumination. When $F^-$ reaches its highest value $1/2$, the two states are completely indistinguishable and thus the scheme fails. The smaller the value of $F^-$, the higher the resolution of the scheme.

A compact form of quantum fidelity is analytically obtainable when $\rho_{1/2}(t)$ are Gaussian states. Being the two-mode squeezed vacuum state, the initial state $|\psi(0)\rangle$ is Gaussian. This Gaussianity is inherited by $\rho(t)$ due to the quadratic-operator characteristic of $\hat{H}_{\text{s}}$. It ensures that both $\rho_0(t)$ and $\rho_1(t)$ are Gaussian too. Such Gaussian states are fully characterized by the mean vectors ${\bf d}_j$ and the covariance matrices ${\pmb\sigma}_j$ ($j=0,1$), whose elements are defined as $d_{j,l}=\text{Tr}(\hat{R}_{l}\rho_j)$ and $\sigma_{j,lm}=\frac{1}{2}\text{Tr}[(\Delta \hat{R}_{j,l}\Delta \hat{R}_{j,m}+\Delta \hat{R}_{j,m}\Delta \hat{R}_{j,l})\rho_j]$, respectively~\cite{RevModPhys.84.621}. Here, $\Delta\hat{R}_{j,l}=\hat{R}_{l}-d_{j,l}$ and $\hat{\bf R}=(\hat{x}_{\text{A}},\hat{p}_{\text{A}},\hat{x}_{\text{B}},\hat{p}_{\text{B}})^{\text{T}}$, with $\hat{x}_{i}=(\hat{a}_{i}^{\dagger}+\hat{a}_{i})/\sqrt{2}$ and $\hat{p}_{i}=i(\hat{a}_{i}^{\dagger}-\hat{a}_{i})/\sqrt{2}$ being the position and momentum operators. The commutation relations of the elements of $\hat{\bf R}$ are $[\hat{R}_{l},\hat{R}_{m}]=i{\Omega}_{lm}$, which defines a symplectic matrix ${\pmb\Omega}=\Big(
\begin{array}{cc}
0 & 1 \\-1 & 0 \\\end{array}
\Big)^{\oplus 2}$. The quantum fidelity between the Gaussian states $\rho_{0}(t)$ and $\rho_1(t)$ is calculated as~\cite{PhysRevA.86.022340,PhysRevA.93.052330}
\begin{equation}\label{eq:eq5}
F(\rho_{0},\rho_{1})=\frac{\exp[-({\bf d}_{1}-{\bf d}_{0})^{\text{T}}{({\pmb\sigma}_{1}+{\pmb\sigma}_{0})^{-1}\over 2}({\bf d}_{1}-{\bf d}_{0})]}{\sqrt{A}+\sqrt{B}-[(\sqrt{A}+\sqrt{B})^{2}-\Lambda]^{1/2}},
\end{equation}
where $\Lambda=\text{det}(\pmb\sigma_{1}+\pmb\sigma_{0})$, $A=16\text{det}({\pmb\Omega}{\pmb\sigma}_{1}{\pmb\Omega}{\pmb\sigma}_{0}-{\bf 1}_{4}/4)$, and $B=16\text{det}({\pmb\sigma}_{1}+i{\pmb\Omega}/2)\text{det}({\pmb\sigma}_{0}+i{\pmb\Omega}/2)$. It is straightforwardly derived that ${\bf d}_{0/1}=0$, ${\pmb\sigma}_{0}=\text{diag}[{2\bar{n}+1\over 2}{\bf 1}_2,{\cosh(2r)\over 2}{\bf 1}_2]$, and ${\pmb\sigma}_{1}=\xi{\pmb\sigma}+(1-\xi){\pmb\sigma}_{0}$, where $\bar{n}\equiv (e^{\beta\omega_{0}}-1)^{-1}$ being the mean photon number in thermal equilibrium, ${\pmb\sigma}=\left[
\begin{array}{cc}
{\cosh(2r)\over 2}{\bf 1}_2 & \frac{\sinh(2r)}{2}{\bf b} \\ \frac{\sinh(2r)}{2}{\bf b} & {\cosh(2r)\over 2}{\bf 1}_2 \\\end{array}
\right]$, and ${\bf b}=\left[\begin{array}{cc}
-\cos(2\omega_{0}t) & \sin(2\omega_{0}t) \\ \sin(2\omega_{0}t) & \cos(2\omega_{0}t) \\\end{array}\right]$. By expanding Eq. \eqref{eq:eq5} in the power of $\xi$, the leading-order term is given by
\begin{eqnarray}
F^{-}_{\text{ideal}}(t)\simeq& [1-\xi\sqrt{\Theta_{\text{ideal}}}]/2,\label{eq:eq9}
\end{eqnarray}
where
\begin{eqnarray}
\Theta_{\text{ideal}}={(\sinh^{2}r-\bar{n})^{2}\over4\varkappa_{1}}+\frac{4^{-1}\sinh^{2}(2r)}{1+\bar{n}+\varkappa_{2}\sinh^{2}r},\label{idathet}
\end{eqnarray}
with $\varkappa_{1}=\bar{n}(\bar{n}+1)$ and $\varkappa_{2}=2\bar{n}+1$. Equation \eqref {idathet} reveals that $F^-_{\text{ideal}}$ is time-independent. Thus, the optimization of transmission time is not required in the practical realization of the ideal scheme. We plot in Fig.~\ref{fig:fig1}(b) $F^-_{\text{ideal}}$ in different values of $r$ with fixed temperature $\beta$ and reflectivity $\xi$. It is obvious to see that $F^-_{\text{ideal}}$ is efficiently reduced by improving the squeezing parameter $r$. It confirms that the quantum entanglement contained in the initial two-mode squeezed vacuum state indeed can be employed as a resource to enhance the resolution of quantum illumination, which is in good agreement with the previous studies~\cite{doi:10.1126/science.1160627,PhysRevLett.101.253601,Shapiro_2009,Zhuang2023ER,GALLEGOTORROME2024100497,Karsa_2024}.

\section{Noisy situation}\label{sec:sec3}
To check the tolerance of the scheme to the quantum noises, we next investigate that each of the light modes A and B suffers from a local decoherence. It is caused by the inevitable interactions of each mode with a local quantum noise during the transmission process [see Fig.~\ref{fig:fig1}(a)]. The Hamiltonian of the total system consisting of the two light beams and the quantum noises reads $\hat{H}_{\text{tot}}=\sum_{l=\text{A},\text{B}}\hat{H}_{l}$, with
\begin{equation}
\hat{H}_{l}=\omega_0\hat{a}_{l}^{\dagger}\hat{a}_{l}+\sum_{k}\omega_{lk}\hat{b}_{lk}^{\dagger}\hat{b}_{lk}+\sum_{k}(g_{lk}\hat{a}_{l}^{\dagger}\hat{b}_{lk}+\text{H}.\text{c}.),
\end{equation}
where $\hat{b}_{lk}$ are the annihilation operators of the $k$th mode of the $l$th quantum noise with frequency $\omega_{lk}$ and $g_{lk}$ is the coupling strength of the $k$th noise mode to the $l$th light mode. The coupling strength is further characterized by the spectral density defined as $J_{l}(\omega)\equiv\sum_{k}|g_{lk}|^{2}\delta(\omega-\omega_{lk})$. We assume that the spectral densities of the two quantum noises have the same Ohmic-family form $J(\omega)=\eta\omega^{s}\omega_{c}^{1-s}e^{-\omega/\omega_{c}}$, where $\eta$ is a dimensionless coupling constant, $\omega_{c}$ is a cutoff frequency, and $s$ is an Ohmicity index. Depending on the value of $s$, the quantum noise is classified into sub-Ohmic when $0<s<1$, Ohmic when $s=1$, and super-Ohmic when $s>1$ \cite{RevModPhys.59.1}. Using the Feynman-Vernon's influence functional method in the coherent-state representation under the condition that the two quantum noises are in the vacuum state initially~\cite{PhysRevA.76.042127}, we exactly trace over the degrees of freedom of the quantum noises and derive a non-Markovian master equation for the two light beams as~\cite{An_2009,PhysRevA.88.012129,SupplementalMaterial}
\begin{equation}\label{eq:eq12}
\dot{\rho}(t)=\sum_{l=\text{A},\text{B}}\big\{-i\varpi(t)[\hat{a}_l^{\dagger}\hat{a}_{l},\rho(t)]+\gamma(t)\mathcal{\check{L}}_{\hat{a}_{l}}\rho(t)\big\},
\end{equation}
where $\mathcal{\check{L}}_{\hat{o}}\cdot\equiv 2\hat{o}\cdot \hat{o}^{\dagger}-\cdot\hat{o}^{\dagger}\hat{o}-\hat{o}^{\dagger}\hat{o}\cdot$ is the Lindblad superoperator, $\varpi(t)=-\text{Im}[\dot{u}(t)/u(t)]$ and $\gamma(t)=-\text{Re}[\dot{u}(t)/u(t)]$ are the time-dependent renormalized frequency and dissipation rate. The coefficient $u(t)$ acts as a decoherence factor and is determined by
\begin{eqnarray}
\dot{u}(t)+i\omega_{0}u(t)+\int_{0}^{t}d\tau\mu(t-\tau)u(\tau)=0,\label{uead}
\end{eqnarray}
with $u(0)=1$ and $\mu(x)=\int_{0}^{\infty}d\omega J(\omega)e^{-i\omega x}$. The convolution in Eqs. \eqref{uead} renders the dynamics non-Markovian, with all the non-Markovian effects being self-consistently incorporated in the coefficients $\varpi(t)$ and $\gamma(t)$.

Solving the master equation \eqref{eq:eq12} under the initial state $|\psi(0)\rangle$, we obtain $\rho(t)$ and its mean vector ${\bf d}=0$ and covariance matrix ${\pmb \sigma}=\left[
\begin{array}{cc}
\frac{2{n}(t)+1}{2}{\bf 1}_2 & \frac{\sinh(2r)}{2}{\bf b} \\ \frac{\sinh(2r)}{2}{\bf b} & \frac{2{n}(t)+1}{2}{\bf 1}_2 \\\end{array}
\right]$, with ${\bf b}=\left[
\begin{array}{cc}
-\text{Re}u(t)^2 & -\text{Im}u(t)^2 \\- \text{Im}u(t)^2& \text{Re}u(t)^2 \\\end{array}
\right]$ and ${n}(t)=|u(t)|^{2}\sinh^{2}r$~\cite{SupplementalMaterial}. Then the mean vectors and the covariance matrices of $\rho_{0/1}(t)$ are calculated to be ${\bf d}_{0/1}=0$, ${\pmb\sigma}_{0}=\text{diag}[\frac{2\bar{n}+1}{2}{\bf 1}_2,\frac{2{n}(t)+1}{2}{\bf 1}_2]$, and ${\pmb\sigma}_{1}=\xi{\pmb \sigma}+(1-\xi){\pmb\sigma}_{0}$. Using Eq.~(\ref{eq:eq5}), we derive
\begin{eqnarray}
F^{-}(t)\simeq [1-\xi\sqrt{\Theta(t)}]/2,
\end{eqnarray} where
\begin{eqnarray}
\Theta(t)=\frac{1+4\varkappa_{1}[1+8\sinh^2(2r)|u(t)|^{4}]+\sum_{j}\lambda_{j}\bar{N}^{j}}{16\varkappa_{1}(1+\varkappa_{2}\bar{N})},~~\label{nctet}
\end{eqnarray}
with $\bar{N}=2n(t)+1$, $\lambda_{1}=2\bar{n}(4\varkappa_{1}+\varkappa_{2})-1$, $\lambda_{2}=-1-8\varkappa_{1}$, and $\lambda_{3}=\varkappa_{2}$.
In the ideal limit, we have $u_{\text{ideal}}(t)=e^{-i\omega_{0}t}$, from which Eq.~\eqref{nctet} reduces to Eq.~\eqref{idathet}.

In the special case when the coupling between each light mode and its quantum noise is weak and the correlation time scale of the noise is smaller than that of the light mode, we can safely apply the Born-Markov approximation to Eq. \eqref{uead} and obtain~\cite{PhysRevE.90.022122} $u_\text{BMA}(t)= \exp\{-\kappa t-i[\omega_0+\Delta(\omega_0)]t\}$. Here, the decay rate $\kappa=\pi J(\omega_0)$ and the frequency shift $\Delta(\omega_0)=\mathcal{P}\int _0^\infty {J(\omega)\over \omega_0-\omega}d\omega$, with $\mathcal{P}$ denoting the Cauchy principal value, reduce to time-independent constants. The approximation leads to $n_{\text{BMA}}(\infty)=0$. Thus, we easily find $F^{-}_{\text{BMA}}(\infty)\simeq1/2$ irrespective of the value of $r$ when the reflectivity is very small. It implies that $\rho_{0}(t)$ and $\rho_{1}(t)$ are completely indistinguishable in the long-time regime. Figure~\ref{fig:fig1}(b) shows the evolution of $F_\text{BMA}^-(t)$. It confirms that $F_\text{BMA}^-(t)$ exclusively increases to $1/2$. Therefore, the quantum superiority induced by quantum entanglement on the resolution of the quantum illumination is entirely destroyed by the Born-Markov decoherence. This result is consistent with the previous ones in Refs.~\cite{PhysRevA.90.052308,Park2023,PhysRevA.111.022627,Zhang_2025}.

%\section{Non-Markovian case}\label{sec:sec5}
To assess whether such a destruction of the quantum superiority by the noise-induced decoherence is essential in physics or artificially caused by the approximation, we now evaluate the performance by relaxing the Born-Markov approximation. In the general non-Markovian case, the analytical expression of $F^-(t)$ is difficult to obtain. We leave it to the numerical calculations. However, via analyzing the long-time behavior of $u(t)$, we may obtain its asymptotic form. A Laplace transform to Eq.~\eqref{uead} results in $\tilde{u}(z)=[z+i\omega_0+\int_0^\infty{J(\omega)\over z+i\omega}d\omega]^{-1}$. Then $u(t)$ is obtained by making the inverse Laplace transform to $\tilde{u}(z)$, which requires finding its pole from
\begin{equation}\label{eq:eq19}
\bar{y}(E)\equiv\omega_0-\int_0^\infty{J(\omega)\over\omega-E}d\omega =E,
\end{equation}
where $E=iz$. It is interesting to discover that the roots $E$ of Eq.~(\ref{eq:eq19}) is just the eigenenergies of $\hat{H}_{l}$ in the single-excitation subspace. Specifically, by expanding the eigenstate as $|\Psi\rangle=(c\hat{a}_{l}^{\dagger}+\sum_{k}d_{lk}\hat{b}_{lk}^{\dagger})|0_{l},\{0_{lk}\}\rangle$ and substituting it into $\hat{H}_{l}|\Psi\rangle=E|\Psi\rangle$ with $E$ being the eigenenergy, we have $(E-\omega_{0})c=\sum_{lk}g_{lk}d_{lk}$ and $d_{lk}=g_{lk}c/(E-\omega_{lk})$. They readily lead to Eq.~(\ref{eq:eq19}) in the continuous limit of the frequencies of the quantum noise. It reveals that, although the subspaces with any excitation numbers are involved, the dynamics governed by $u(t)$ is uniquely determined by the energy-spectrum characteristic of $\hat{H}_{l}$ in the single-excitation subspace. This supplies us an insightful guideline to control the decoherence dynamics of the light modes via engineering the energy-spectrum feature. Since $\bar{y}(E)$ is a decreasing function in the regime $E < 0$, Eq.~(\ref{eq:eq19}) has one isolated root $E_b$ in this regime provided $\bar{y}(0) < 0$.  We call the eigenstate of the isolated eigenenergy $E_b$ bound state. In the regime $E>0$, $\bar{y}(E)$ is divergent and thus Eq.~(\ref{eq:eq19}) has infinite number of roots forming a continuous energy band. After making the inverse Laplace transform to $\tilde{u}(z)$, we obtain~\cite{PhysRevA.103.L010601}
\begin{equation}
u(t)=Ze^{-iE_b t}+\int_{0}^{\infty}dE\tfrac{J(E)e^{-iE t}}{[E-\omega_{0}-\Delta(E)]^{2}+[\pi J(E)]^2},
\end{equation}
where the first term with $Z=[1+\int_0^\infty{J(\omega)d\omega\over(E_b-\omega)^2}]^{-1}$ is the residue contributed by the bound state and the second is the branch cut contributed by the band energies. Oscillating with time in continuously changing frequencies, the integral tends to zero in the long-time condition due to out-of-phase interference. Thus, if the bound state is absent, then $u(\infty)= 0$ characterizes a complete decoherence, while if the bound state is formed, then $u(\infty)=Ze^{-iE_b t}$ implies a decoherence suppression. It is easy to evaluate from $\bar{y}(0)<0$ that the bound state for the Ohmic-family spectral density is formed when $\omega_0< \eta\omega_c \underline{\Gamma}(s)$, where $\underline{\Gamma}(s)$ is the Euler's gamma function.

\begin{figure}
\centering
\includegraphics[angle=0,width=0.47\textwidth]{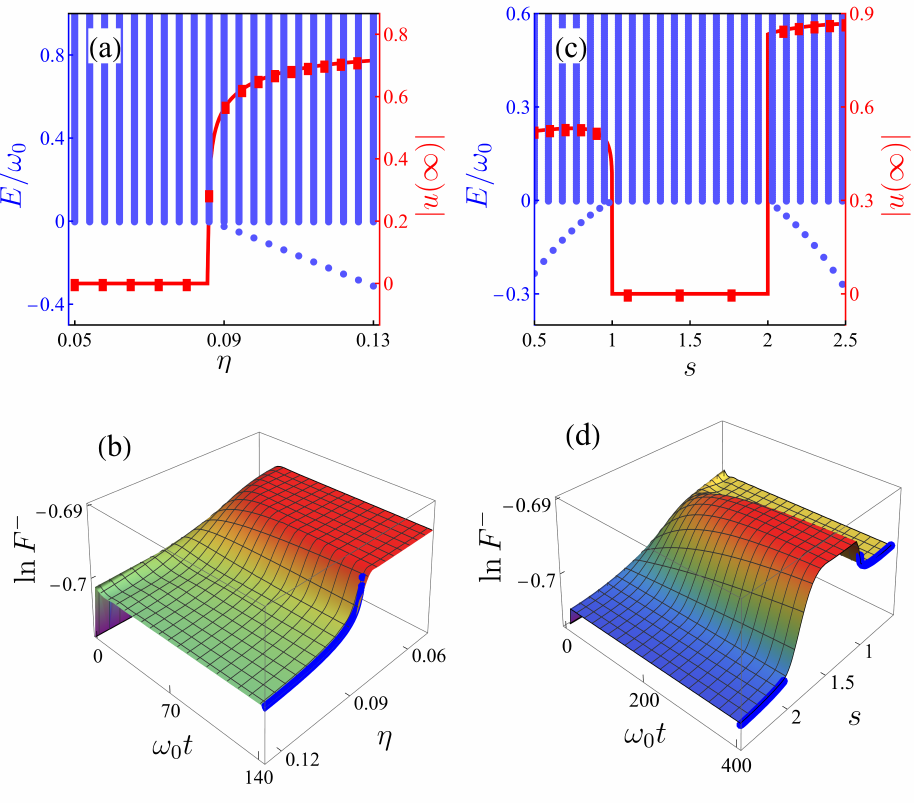}
\caption{(a) Energy spectrum (blue dots) via solving Eq. \eqref{eq:eq19}, $|u(t=400\omega_{0}^{-1})|$ (red rectangles) via solving Eq. \eqref{uead}, and $Z$ (red lines) via evaluating the the residue contributed by the bound state as a function of $\eta$ when $s=0.8$ and $\omega_{c}=10\omega_{0}$ in (a) and $s$ when $\eta=0.2$ and $\omega_{c}=5\omega_0$ in (c). Evolution of the corresponding $F^{-}(t)$ in different values of (b) $\eta$ and (d) $s$. The analytical results predicted by Eq. (\ref{eq:eq21}) in the presence of the bound state are presented by the blue lines. Other parameters are $\xi=10^{-3}$ and $\beta=2\omega_{0}^{-1}$.}\label{fig:fig2}
\end{figure}

It is natural to expect that $F^-(t)$ has the same behavior as that of the Born-Markov approximate result when the bound state is absent. Therefore, we pay our attention to the case in the presence of the bound state. It is surprising to find that, when the bound state is present, the long-time form of $\Theta(t)$ becomes
\begin{equation}\label{eq:eq21}
\Theta(\infty)={(Z^{2}\sinh^{2}r-\bar{n})^{2}\over4\varkappa_{1}}+\frac{4^{-1}Z^{4}\sinh^{2}(2r)}{1+\bar{n}+\varkappa_{2}Z^{2}\sinh^{2}r}.
\end{equation}
It is obvious to see that $\Theta(\infty)$ sensitively depends on the values of $Z$ and $r$, which is in sharp contrast to that of the Born-Markov approximate case. We thus achieve a higher steady-state resolution in the non-Markovian dynamics due to the formation of the bound state. On the one hand, the value of $Z$ is tunable via changing the values of $\omega_{0}$ or the parameters of the spectral density. If we manage to tune $Z$ as close as possible to 1, we would retrieve the ideal form of $\Theta_{\text{ideal}}$ in Eq. \eqref{idathet}. This result implies that the ideal performance of the quantum illumination is asymptotically recovered in the non-Markovian dynamics with the help of the bound state. On the other hand, the formation of the bound state partially preserves the initial quantum entanglement in the steady state, which enables us to improve the resolution by increasing $r$. Therefore, the quantum superiority, which is completely destroyed by the decoherence under the Born-Markov approximation, is retrieved in the non-Markovian dynamics due to the formation of the bound state.

\begin{figure}
\centering
\includegraphics[angle=0,width=0.47\textwidth]{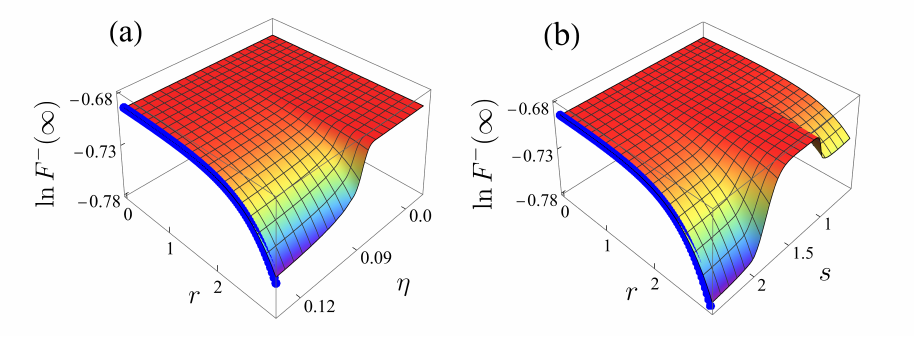}
\caption{Steady-state $F^-$ at $t=400\omega_0^{-1}$ in different values of the squeezing parameter $r$ as the functions of (a) $\eta$ and (b) $s$. The blue lines denote the analytical results predicted by Eq. (\ref{eq:eq21}). Other parameters are the same as Fig.~\ref{fig:fig2}.}\label{fig:fig3}
\end{figure}

We now perform the numerical calculations to verify our results. Figure~\ref{fig:fig2}(a) shows the energy spectrum of $\hat{H}_{l}$ in the single-excitation subspace as a function of $\eta$. It is found that a bound state is present when $\eta>0.086$, which matches our analytical criteria $\eta\geq\omega_{0}/[\omega_c \underline{\Gamma}(s)]$. With the formation of the bound state, the decoherence factor $|u(t)|$ evolves exclusively to finite values matching exactly those analytically evaluated from $Z$. The evolution of the corresponding $F^-$ is displayed in Fig.~\ref{fig:fig2}(b). An obvious threshold of $F^{-}$ is observed at the critical point of forming the bound state. When the bound state is absent, $F^{-}$ exclusively tends to $1/2$ in the long-time regime, which means the capacity of distinguishing $\rho_{0}$ and $\rho_{1}$ as well as the scheme of quantum illumination are completely destroyed by the decoherence. This result is qualitatively similar to the one under the Born-Markov approximation. In contrast, when the bound state is present, we have $F^{-}(\infty)<1/2$, which means that the capacity of distinguishing $\rho_{0}$ and $\rho_{1}$ is recovered. In this case, the quantum superiority of the quantum illumination in the presence of the quantum noises is recovered in the non-Markovian dynamics. Figures \ref{fig:fig2}(c) and \ref{fig:fig2}(d) show the results under different values of the Ohmicity index $s$. The bound state is formed as long as $\underline{\Gamma}(s)>\omega_0/(\eta\omega_c)$. With the formation of the bound state, $F^-(t)$ tends to a value smaller than $1/2$. In particular, with $Z$ approaching 1 for larger $s$, $F^-(\infty)$ becomes smaller and smaller. Figure~\ref{fig:fig3} shows the long-time behaviors of $F^{-}$ with different squeezing parameters. It numerically confirms that, in sharp different from the constant value $1/2$ when the bound state is absent, $F^-(\infty)$ exhibits a decreasing dependence on $r$. Therefore, the constructive role of the quantum squeezing in improving the resolution of quantum illumination is restored.

\section{Discussion and conclusion}\label{sec:sec6}
Our result reveals that we can improve the resolution of noisy quantum illumination via engineering the formation of the bound state, which is realizable via the quantum reservoir engineering technique~\cite{ER1,Kienzler53,PhysRevA.78.010101,Kitzman2023}. It is noted that, although only the Ohmic-family spectral density is considered, our bound-state mechanism to recover the ideal performation of quantum illumination is applicable to other spectral forms, where the explicit condition for forming the bound state may be different, but the bound-state mechanism does not change. This endows our mechanism with universality in suppressing decoherence in quantum technologies. Especially, the bound state and its dynamical effect have been experimentally observed in circuit QED~\cite{Liu2017} and ultracold atom~\cite{Krinner2018,RN11} systems. These experimental progresses provide strong support to test the performance of our noisy quantum illumination scheme in realistic settings.

In summary, we have discovered a mechanism to retrieve the ideal superiority of the quantum illumination under the influence of the practical decoherence induced by local noises. It has been revealed that the resolution of such a noisy quantum illumination is essentially governed by the feature of the energy spectrum of the total system consisting of each light mode and its local quantum noise. With the formation of a bound state in the energy spectrum, the resolution asymptotically returns to its ideal behavior. In contrast to the results in previous works based on the Born-Markov approximation, where the quantum superiority of the quantum illumination is completely destroyed by the decoherence, our result supplies a useful guideline in suppressing the noisy effect of quantum illumination. It lays a firm foundation for realizing a high-resolution quantum radar in the noisy intermediate-scale quantum era.

\section{Acknowledgments}
The work is supported by the National Natural Science Foundation of China (Grants No. 12375015, No. 12275109, and No. 12247101), the Innovation Program for Quantum Science and Technology of China (Grant No. 2023ZD0300904), the Natural Science Foundation of Gansu Province (No. 22JR5RA389 and No. 25JRRA799), and the Fundamental Research Funds for the Central Universities (Grant No. lzujbky-2025-jdzx07), and the ‘111 Center’ (Grant No. B20063).

\bibliography{reference}

\begin{thebibliography}{64}%
\makeatletter
\providecommand \@ifxundefined [1]{%
 \@ifx{#1\undefined}
}%
\providecommand \@ifnum [1]{%
 \ifnum #1\expandafter \@firstoftwo
 \else \expandafter \@secondoftwo
 \fi
}%
\providecommand \@ifx [1]{%
 \ifx #1\expandafter \@firstoftwo
 \else \expandafter \@secondoftwo
 \fi
}%
\providecommand \natexlab [1]{#1}%
\providecommand \enquote  [1]{``#1''}%
\providecommand \bibnamefont  [1]{#1}%
\providecommand \bibfnamefont [1]{#1}%
\providecommand \citenamefont [1]{#1}%
\providecommand \href@noop [0]{\@secondoftwo}%
\providecommand \href [0]{\begingroup \@sanitize@url \@href}%
\providecommand \@href[1]{\@@startlink{#1}\@@href}%
\providecommand \@@href[1]{\endgroup#1\@@endlink}%
\providecommand \@sanitize@url [0]{\catcode `\\12\catcode `\$12\catcode `\&12\catcode `\#12\catcode `\^12\catcode `\_12\catcode `\%12\relax}%
\providecommand \@@startlink[1]{}%
\providecommand \@@endlink[0]{}%
\providecommand \url  [0]{\begingroup\@sanitize@url \@url }%
\providecommand \@url [1]{\endgroup\@href {#1}{\urlprefix }}%
\providecommand \urlprefix  [0]{URL }%
\providecommand \Eprint [0]{\href }%
\providecommand \doibase [0]{https://doi.org/}%
\providecommand \selectlanguage [0]{\@gobble}%
\providecommand \bibinfo  [0]{\@secondoftwo}%
\providecommand \bibfield  [0]{\@secondoftwo}%
\providecommand \translation [1]{[#1]}%
\providecommand \BibitemOpen [0]{}%
\providecommand \bibitemStop [0]{}%
\providecommand \bibitemNoStop [0]{.\EOS\space}%
\providecommand \EOS [0]{\spacefactor3000\relax}%
\providecommand \BibitemShut  [1]{\csname bibitem#1\endcsname}%
\let\auto@bib@innerbib\@empty
%</preamble>
\bibitem [{\citenamefont {MacFarlane}\ \emph {et~al.}(2003)\citenamefont {MacFarlane}, \citenamefont {Dowling},\ and\ \citenamefont {Milburn}}]{doi:10.1098/rsta.2003.1227}%
  \BibitemOpen
  \bibfield  {author} {\bibinfo {author} {\bibfnamefont {A.~G.~J.}\ \bibnamefont {MacFarlane}}, \bibinfo {author} {\bibfnamefont {J.~P.}\ \bibnamefont {Dowling}},\ and\ \bibinfo {author} {\bibfnamefont {G.~J.}\ \bibnamefont {Milburn}},\ }\bibfield  {title} {\bibinfo {title} {Quantum technology: the second quantum revolution},\ }\href {https://doi.org/10.1098/rsta.2003.1227} {\bibfield  {journal} {\bibinfo  {journal} {Philosophical Transactions of the Royal Society of London. Series A: Mathematical, Physical and Engineering Sciences}\ }\textbf {\bibinfo {volume} {361}},\ \bibinfo {pages} {1655} (\bibinfo {year} {2003})}\BibitemShut {NoStop}%
\bibitem [{\citenamefont {Lloyd}(2008)}]{doi:10.1126/science.1160627}%
  \BibitemOpen
  \bibfield  {author} {\bibinfo {author} {\bibfnamefont {S.}~\bibnamefont {Lloyd}},\ }\bibfield  {title} {\bibinfo {title} {Enhanced sensitivity of photodetection via quantum illumination},\ }\href {https://doi.org/10.1126/science.1160627} {\bibfield  {journal} {\bibinfo  {journal} {Science}\ }\textbf {\bibinfo {volume} {321}},\ \bibinfo {pages} {1463} (\bibinfo {year} {2008})}\BibitemShut {NoStop}%
\bibitem [{\citenamefont {Tan}\ \emph {et~al.}(2008)\citenamefont {Tan}, \citenamefont {Erkmen}, \citenamefont {Giovannetti}, \citenamefont {Guha}, \citenamefont {Lloyd}, \citenamefont {Maccone}, \citenamefont {Pirandola},\ and\ \citenamefont {Shapiro}}]{PhysRevLett.101.253601}%
  \BibitemOpen
  \bibfield  {author} {\bibinfo {author} {\bibfnamefont {S.-H.}\ \bibnamefont {Tan}}, \bibinfo {author} {\bibfnamefont {B.~I.}\ \bibnamefont {Erkmen}}, \bibinfo {author} {\bibfnamefont {V.}~\bibnamefont {Giovannetti}}, \bibinfo {author} {\bibfnamefont {S.}~\bibnamefont {Guha}}, \bibinfo {author} {\bibfnamefont {S.}~\bibnamefont {Lloyd}}, \bibinfo {author} {\bibfnamefont {L.}~\bibnamefont {Maccone}}, \bibinfo {author} {\bibfnamefont {S.}~\bibnamefont {Pirandola}},\ and\ \bibinfo {author} {\bibfnamefont {J.~H.}\ \bibnamefont {Shapiro}},\ }\bibfield  {title} {\bibinfo {title} {Quantum illumination with gaussian states},\ }\href {https://doi.org/10.1103/PhysRevLett.101.253601} {\bibfield  {journal} {\bibinfo  {journal} {Phys. Rev. Lett.}\ }\textbf {\bibinfo {volume} {101}},\ \bibinfo {pages} {253601} (\bibinfo {year} {2008})}\BibitemShut {NoStop}%
\bibitem [{\citenamefont {Shapiro}\ and\ \citenamefont {Lloyd}(2009)}]{Shapiro_2009}%
  \BibitemOpen
  \bibfield  {author} {\bibinfo {author} {\bibfnamefont {J.~H.}\ \bibnamefont {Shapiro}}\ and\ \bibinfo {author} {\bibfnamefont {S.}~\bibnamefont {Lloyd}},\ }\bibfield  {title} {\bibinfo {title} {Quantum illumination versus coherent-state target detection},\ }\href {https://doi.org/10.1088/1367-2630/11/6/063045} {\bibfield  {journal} {\bibinfo  {journal} {New Journal of Physics}\ }\textbf {\bibinfo {volume} {11}},\ \bibinfo {pages} {063045} (\bibinfo {year} {2009})}\BibitemShut {NoStop}%
\bibitem [{\citenamefont {Blakey}\ \emph {et~al.}(2022)\citenamefont {Blakey}, \citenamefont {Liu}, \citenamefont {Papangelakis}, \citenamefont {Zhang}, \citenamefont {L{\'e}ger}, \citenamefont {Iu},\ and\ \citenamefont {Helmy}}]{Blakey2022}%
  \BibitemOpen
  \bibfield  {author} {\bibinfo {author} {\bibfnamefont {P.~S.}\ \bibnamefont {Blakey}}, \bibinfo {author} {\bibfnamefont {H.}~\bibnamefont {Liu}}, \bibinfo {author} {\bibfnamefont {G.}~\bibnamefont {Papangelakis}}, \bibinfo {author} {\bibfnamefont {Y.}~\bibnamefont {Zhang}}, \bibinfo {author} {\bibfnamefont {Z.~M.}\ \bibnamefont {L{\'e}ger}}, \bibinfo {author} {\bibfnamefont {M.~L.}\ \bibnamefont {Iu}},\ and\ \bibinfo {author} {\bibfnamefont {A.~S.}\ \bibnamefont {Helmy}},\ }\bibfield  {title} {\bibinfo {title} {Quantum and non-local effects offer over 40{\thinspace}db noise resilience advantage towards quantum lidar},\ }\href {https://doi.org/10.1038/s41467-022-33376-9} {\bibfield  {journal} {\bibinfo  {journal} {Nature Communications}\ }\textbf {\bibinfo {volume} {13}},\ \bibinfo {pages} {5633} (\bibinfo {year} {2022})}\BibitemShut {NoStop}%
\bibitem [{\citenamefont {Liu}\ \emph {et~al.}(2023)\citenamefont {Liu}, \citenamefont {Qin}, \citenamefont {Papangelakis}, \citenamefont {Iu},\ and\ \citenamefont {Helmy}}]{Liu2023}%
  \BibitemOpen
  \bibfield  {author} {\bibinfo {author} {\bibfnamefont {H.}~\bibnamefont {Liu}}, \bibinfo {author} {\bibfnamefont {C.}~\bibnamefont {Qin}}, \bibinfo {author} {\bibfnamefont {G.}~\bibnamefont {Papangelakis}}, \bibinfo {author} {\bibfnamefont {M.~L.}\ \bibnamefont {Iu}},\ and\ \bibinfo {author} {\bibfnamefont {A.~S.}\ \bibnamefont {Helmy}},\ }\bibfield  {title} {\bibinfo {title} {Compact all-fiber quantum-inspired lidar with over 100{\thinspace}db noise rejection and single photon sensitivity},\ }\href {https://doi.org/10.1038/s41467-023-40914-6} {\bibfield  {journal} {\bibinfo  {journal} {Nature Communications}\ }\textbf {\bibinfo {volume} {14}},\ \bibinfo {pages} {5344} (\bibinfo {year} {2023})}\BibitemShut {NoStop}%
\bibitem [{\citenamefont {Dalvit}\ \emph {et~al.}(2024)\citenamefont {Dalvit}, \citenamefont {Volkoff}, \citenamefont {Choi}, \citenamefont {Azad}, \citenamefont {Chen},\ and\ \citenamefont {Milonni}}]{PhysRevX.14.041058}%
  \BibitemOpen
  \bibfield  {author} {\bibinfo {author} {\bibfnamefont {D.~A.~R.}\ \bibnamefont {Dalvit}}, \bibinfo {author} {\bibfnamefont {T.~J.}\ \bibnamefont {Volkoff}}, \bibinfo {author} {\bibfnamefont {Y.-S.}\ \bibnamefont {Choi}}, \bibinfo {author} {\bibfnamefont {A.~K.}\ \bibnamefont {Azad}}, \bibinfo {author} {\bibfnamefont {H.-T.}\ \bibnamefont {Chen}},\ and\ \bibinfo {author} {\bibfnamefont {P.~W.}\ \bibnamefont {Milonni}},\ }\bibfield  {title} {\bibinfo {title} {Quantum frequency combs with path identity for quantum remote sensing},\ }\href {https://doi.org/10.1103/PhysRevX.14.041058} {\bibfield  {journal} {\bibinfo  {journal} {Phys. Rev. X}\ }\textbf {\bibinfo {volume} {14}},\ \bibinfo {pages} {041058} (\bibinfo {year} {2024})}\BibitemShut {NoStop}%
\bibitem [{\citenamefont {Kornienko}\ \emph {et~al.}(2024)\citenamefont {Kornienko}, \citenamefont {Vidal}, \citenamefont {Pönni}, \citenamefont {Raasakka},\ and\ \citenamefont {Tittonen}}]{10.1063/5.0162419}%
  \BibitemOpen
  \bibfield  {author} {\bibinfo {author} {\bibfnamefont {V.~V.}\ \bibnamefont {Kornienko}}, \bibinfo {author} {\bibfnamefont {C.}~\bibnamefont {Vidal}}, \bibinfo {author} {\bibfnamefont {A.}~\bibnamefont {Pönni}}, \bibinfo {author} {\bibfnamefont {M.}~\bibnamefont {Raasakka}},\ and\ \bibinfo {author} {\bibfnamefont {I.}~\bibnamefont {Tittonen}},\ }\bibfield  {title} {\bibinfo {title} {Partially reflecting jamming objects in correlation-enhanced target detection with entangled photons},\ }\href {https://doi.org/10.1063/5.0162419} {\bibfield  {journal} {\bibinfo  {journal} {APL Quantum}\ }\textbf {\bibinfo {volume} {1}},\ \bibinfo {pages} {016107} (\bibinfo {year} {2024})}\BibitemShut {NoStop}%
\bibitem [{\citenamefont {Gregory}\ \emph {et~al.}(2020)\citenamefont {Gregory}, \citenamefont {Moreau}, \citenamefont {Toninelli},\ and\ \citenamefont {Padgett}}]{doi:10.1126/sciadv.aay2652}%
  \BibitemOpen
  \bibfield  {author} {\bibinfo {author} {\bibfnamefont {T.}~\bibnamefont {Gregory}}, \bibinfo {author} {\bibfnamefont {P.-A.}\ \bibnamefont {Moreau}}, \bibinfo {author} {\bibfnamefont {E.}~\bibnamefont {Toninelli}},\ and\ \bibinfo {author} {\bibfnamefont {M.~J.}\ \bibnamefont {Padgett}},\ }\bibfield  {title} {\bibinfo {title} {Imaging through noise with quantum illumination},\ }\href {https://doi.org/10.1126/sciadv.aay2652} {\bibfield  {journal} {\bibinfo  {journal} {Science Advances}\ }\textbf {\bibinfo {volume} {6}},\ \bibinfo {pages} {eaay2652} (\bibinfo {year} {2020})}\BibitemShut {NoStop}%
\bibitem [{\citenamefont {Sanz}\ \emph {et~al.}(2017)\citenamefont {Sanz}, \citenamefont {Las~Heras}, \citenamefont {Garc\'{\i}a-Ripoll}, \citenamefont {Solano},\ and\ \citenamefont {Di~Candia}}]{PhysRevLett.118.070803}%
  \BibitemOpen
  \bibfield  {author} {\bibinfo {author} {\bibfnamefont {M.}~\bibnamefont {Sanz}}, \bibinfo {author} {\bibfnamefont {U.}~\bibnamefont {Las~Heras}}, \bibinfo {author} {\bibfnamefont {J.~J.}\ \bibnamefont {Garc\'{\i}a-Ripoll}}, \bibinfo {author} {\bibfnamefont {E.}~\bibnamefont {Solano}},\ and\ \bibinfo {author} {\bibfnamefont {R.}~\bibnamefont {Di~Candia}},\ }\bibfield  {title} {\bibinfo {title} {Quantum estimation methods for quantum illumination},\ }\href {https://doi.org/10.1103/PhysRevLett.118.070803} {\bibfield  {journal} {\bibinfo  {journal} {Phys. Rev. Lett.}\ }\textbf {\bibinfo {volume} {118}},\ \bibinfo {pages} {070803} (\bibinfo {year} {2017})}\BibitemShut {NoStop}%
\bibitem [{\citenamefont {Michael}\ \emph {et~al.}(2021)\citenamefont {Michael}, \citenamefont {Jonas}, \citenamefont {Bello}, \citenamefont {Meller}, \citenamefont {Cohen}, \citenamefont {Rosenbluh},\ and\ \citenamefont {Pe'er}}]{PhysRevLett.127.173603}%
  \BibitemOpen
  \bibfield  {author} {\bibinfo {author} {\bibfnamefont {Y.}~\bibnamefont {Michael}}, \bibinfo {author} {\bibfnamefont {I.}~\bibnamefont {Jonas}}, \bibinfo {author} {\bibfnamefont {L.}~\bibnamefont {Bello}}, \bibinfo {author} {\bibfnamefont {M.-E.}\ \bibnamefont {Meller}}, \bibinfo {author} {\bibfnamefont {E.}~\bibnamefont {Cohen}}, \bibinfo {author} {\bibfnamefont {M.}~\bibnamefont {Rosenbluh}},\ and\ \bibinfo {author} {\bibfnamefont {A.}~\bibnamefont {Pe'er}},\ }\bibfield  {title} {\bibinfo {title} {Augmenting the sensing performance of entangled photon pairs through asymmetry},\ }\href {https://doi.org/10.1103/PhysRevLett.127.173603} {\bibfield  {journal} {\bibinfo  {journal} {Phys. Rev. Lett.}\ }\textbf {\bibinfo {volume} {127}},\ \bibinfo {pages} {173603} (\bibinfo {year} {2021})}\BibitemShut {NoStop}%
\bibitem [{\citenamefont {Xu}\ \emph {et~al.}(2021)\citenamefont {Xu}, \citenamefont {Zhang}, \citenamefont {Xu}, \citenamefont {Jiang}, \citenamefont {Yung},\ and\ \citenamefont {Zhang}}]{PhysRevLett.127.040504}%
  \BibitemOpen
  \bibfield  {author} {\bibinfo {author} {\bibfnamefont {F.}~\bibnamefont {Xu}}, \bibinfo {author} {\bibfnamefont {X.-M.}\ \bibnamefont {Zhang}}, \bibinfo {author} {\bibfnamefont {L.}~\bibnamefont {Xu}}, \bibinfo {author} {\bibfnamefont {T.}~\bibnamefont {Jiang}}, \bibinfo {author} {\bibfnamefont {M.-H.}\ \bibnamefont {Yung}},\ and\ \bibinfo {author} {\bibfnamefont {L.}~\bibnamefont {Zhang}},\ }\bibfield  {title} {\bibinfo {title} {Experimental quantum target detection approaching the fundamental helstrom limit},\ }\href {https://doi.org/10.1103/PhysRevLett.127.040504} {\bibfield  {journal} {\bibinfo  {journal} {Phys. Rev. Lett.}\ }\textbf {\bibinfo {volume} {127}},\ \bibinfo {pages} {040504} (\bibinfo {year} {2021})}\BibitemShut {NoStop}%
\bibitem [{\citenamefont {Qian}\ \emph {et~al.}(2023)\citenamefont {Qian}, \citenamefont {Xu}, \citenamefont {Zhu}, \citenamefont {Xu}, \citenamefont {Gao}, \citenamefont {Yakovlev}, \citenamefont {Liu}, \citenamefont {Zhu},\ and\ \citenamefont {Wang}}]{PhysRevLett.131.033603}%
  \BibitemOpen
  \bibfield  {author} {\bibinfo {author} {\bibfnamefont {G.}~\bibnamefont {Qian}}, \bibinfo {author} {\bibfnamefont {X.}~\bibnamefont {Xu}}, \bibinfo {author} {\bibfnamefont {S.-A.}\ \bibnamefont {Zhu}}, \bibinfo {author} {\bibfnamefont {C.}~\bibnamefont {Xu}}, \bibinfo {author} {\bibfnamefont {F.}~\bibnamefont {Gao}}, \bibinfo {author} {\bibfnamefont {V.~V.}\ \bibnamefont {Yakovlev}}, \bibinfo {author} {\bibfnamefont {X.}~\bibnamefont {Liu}}, \bibinfo {author} {\bibfnamefont {S.-Y.}\ \bibnamefont {Zhu}},\ and\ \bibinfo {author} {\bibfnamefont {D.-W.}\ \bibnamefont {Wang}},\ }\bibfield  {title} {\bibinfo {title} {Quantum induced coherence light detection and ranging},\ }\href {https://doi.org/10.1103/PhysRevLett.131.033603} {\bibfield  {journal} {\bibinfo  {journal} {Phys. Rev. Lett.}\ }\textbf {\bibinfo {volume} {131}},\ \bibinfo {pages} {033603} (\bibinfo {year} {2023})}\BibitemShut {NoStop}%
\bibitem [{\citenamefont {Volkoff}(2024)}]{Volkoff_2024}%
  \BibitemOpen
  \bibfield  {author} {\bibinfo {author} {\bibfnamefont {T.~J.}\ \bibnamefont {Volkoff}},\ }\bibfield  {title} {\bibinfo {title} {Not even 6 db: Gaussian quantum illumination in thermal background},\ }\href {https://doi.org/10.1088/1751-8121/ad1e18} {\bibfield  {journal} {\bibinfo  {journal} {Journal of Physics A: Mathematical and Theoretical}\ }\textbf {\bibinfo {volume} {57}},\ \bibinfo {pages} {065301} (\bibinfo {year} {2024})}\BibitemShut {NoStop}%
\bibitem [{\citenamefont {Zhuang}\ \emph {et~al.}(2017)\citenamefont {Zhuang}, \citenamefont {Zhang},\ and\ \citenamefont {Shapiro}}]{PhysRevLett.118.040801}%
  \BibitemOpen
  \bibfield  {author} {\bibinfo {author} {\bibfnamefont {Q.}~\bibnamefont {Zhuang}}, \bibinfo {author} {\bibfnamefont {Z.}~\bibnamefont {Zhang}},\ and\ \bibinfo {author} {\bibfnamefont {J.~H.}\ \bibnamefont {Shapiro}},\ }\bibfield  {title} {\bibinfo {title} {Optimum mixed-state discrimination for noisy entanglement-enhanced sensing},\ }\href {https://doi.org/10.1103/PhysRevLett.118.040801} {\bibfield  {journal} {\bibinfo  {journal} {Phys. Rev. Lett.}\ }\textbf {\bibinfo {volume} {118}},\ \bibinfo {pages} {040801} (\bibinfo {year} {2017})}\BibitemShut {NoStop}%
\bibitem [{\citenamefont {Angeletti}\ \emph {et~al.}(2023)\citenamefont {Angeletti}, \citenamefont {Shi}, \citenamefont {Lakshmanan}, \citenamefont {Vitali},\ and\ \citenamefont {Zhuang}}]{PhysRevApplied.20.024030}%
  \BibitemOpen
  \bibfield  {author} {\bibinfo {author} {\bibfnamefont {J.}~\bibnamefont {Angeletti}}, \bibinfo {author} {\bibfnamefont {H.}~\bibnamefont {Shi}}, \bibinfo {author} {\bibfnamefont {T.}~\bibnamefont {Lakshmanan}}, \bibinfo {author} {\bibfnamefont {D.}~\bibnamefont {Vitali}},\ and\ \bibinfo {author} {\bibfnamefont {Q.}~\bibnamefont {Zhuang}},\ }\bibfield  {title} {\bibinfo {title} {Microwave quantum illumination with correlation-to-displacement conversion},\ }\href {https://doi.org/10.1103/PhysRevApplied.20.024030} {\bibfield  {journal} {\bibinfo  {journal} {Phys. Rev. Appl.}\ }\textbf {\bibinfo {volume} {20}},\ \bibinfo {pages} {024030} (\bibinfo {year} {2023})}\BibitemShut {NoStop}%
\bibitem [{\citenamefont {Chen}\ and\ \citenamefont {Zhuang}(2023)}]{PhysRevA.107.062405}%
  \BibitemOpen
  \bibfield  {author} {\bibinfo {author} {\bibfnamefont {X.}~\bibnamefont {Chen}}\ and\ \bibinfo {author} {\bibfnamefont {Q.}~\bibnamefont {Zhuang}},\ }\bibfield  {title} {\bibinfo {title} {Entanglement-assisted detection of fading targets via correlation-to-displacement conversion},\ }\href {https://doi.org/10.1103/PhysRevA.107.062405} {\bibfield  {journal} {\bibinfo  {journal} {Phys. Rev. A}\ }\textbf {\bibinfo {volume} {107}},\ \bibinfo {pages} {062405} (\bibinfo {year} {2023})}\BibitemShut {NoStop}%
\bibitem [{\citenamefont {Reichert}\ \emph {et~al.}(2024)\citenamefont {Reichert}, \citenamefont {Zhuang},\ and\ \citenamefont {Sanz}}]{PhysRevLett.133.130801}%
  \BibitemOpen
  \bibfield  {author} {\bibinfo {author} {\bibfnamefont {M.}~\bibnamefont {Reichert}}, \bibinfo {author} {\bibfnamefont {Q.}~\bibnamefont {Zhuang}},\ and\ \bibinfo {author} {\bibfnamefont {M.}~\bibnamefont {Sanz}},\ }\bibfield  {title} {\bibinfo {title} {Heisenberg-limited quantum lidar for joint range and velocity estimation},\ }\href {https://doi.org/10.1103/PhysRevLett.133.130801} {\bibfield  {journal} {\bibinfo  {journal} {Phys. Rev. Lett.}\ }\textbf {\bibinfo {volume} {133}},\ \bibinfo {pages} {130801} (\bibinfo {year} {2024})}\BibitemShut {NoStop}%
\bibitem [{\citenamefont {Wilde}\ \emph {et~al.}(2017)\citenamefont {Wilde}, \citenamefont {Tomamichel}, \citenamefont {Lloyd},\ and\ \citenamefont {Berta}}]{PhysRevLett.119.120501}%
  \BibitemOpen
  \bibfield  {author} {\bibinfo {author} {\bibfnamefont {M.~M.}\ \bibnamefont {Wilde}}, \bibinfo {author} {\bibfnamefont {M.}~\bibnamefont {Tomamichel}}, \bibinfo {author} {\bibfnamefont {S.}~\bibnamefont {Lloyd}},\ and\ \bibinfo {author} {\bibfnamefont {M.}~\bibnamefont {Berta}},\ }\bibfield  {title} {\bibinfo {title} {Gaussian hypothesis testing and quantum illumination},\ }\href {https://doi.org/10.1103/PhysRevLett.119.120501} {\bibfield  {journal} {\bibinfo  {journal} {Phys. Rev. Lett.}\ }\textbf {\bibinfo {volume} {119}},\ \bibinfo {pages} {120501} (\bibinfo {year} {2017})}\BibitemShut {NoStop}%
\bibitem [{\citenamefont {Li}\ and\ \citenamefont {Ren}(2023)}]{PhysRevA.108.062605}%
  \BibitemOpen
  \bibfield  {author} {\bibinfo {author} {\bibfnamefont {Y.}~\bibnamefont {Li}}\ and\ \bibinfo {author} {\bibfnamefont {C.}~\bibnamefont {Ren}},\ }\bibfield  {title} {\bibinfo {title} {Entanglement-enhanced quantum strategies for accurate estimation of multibody-group motion and moving-object characteristics},\ }\href {https://doi.org/10.1103/PhysRevA.108.062605} {\bibfield  {journal} {\bibinfo  {journal} {Phys. Rev. A}\ }\textbf {\bibinfo {volume} {108}},\ \bibinfo {pages} {062605} (\bibinfo {year} {2023})}\BibitemShut {NoStop}%
\bibitem [{\citenamefont {Zhang}\ \emph {et~al.}(2024)\citenamefont {Zhang}, \citenamefont {Xia}, \citenamefont {Ye}, \citenamefont {Chang},\ and\ \citenamefont {Liao}}]{PhysRevA.109.062440}%
  \BibitemOpen
  \bibfield  {author} {\bibinfo {author} {\bibfnamefont {H.}~\bibnamefont {Zhang}}, \bibinfo {author} {\bibfnamefont {Y.}~\bibnamefont {Xia}}, \bibinfo {author} {\bibfnamefont {W.}~\bibnamefont {Ye}}, \bibinfo {author} {\bibfnamefont {S.}~\bibnamefont {Chang}},\ and\ \bibinfo {author} {\bibfnamefont {Z.}~\bibnamefont {Liao}},\ }\bibfield  {title} {\bibinfo {title} {Quantum illumination using non-gaussian states with conditional measurements},\ }\href {https://doi.org/10.1103/PhysRevA.109.062440} {\bibfield  {journal} {\bibinfo  {journal} {Phys. Rev. A}\ }\textbf {\bibinfo {volume} {109}},\ \bibinfo {pages} {062440} (\bibinfo {year} {2024})}\BibitemShut {NoStop}%
\bibitem [{\citenamefont {Barzanjeh}\ \emph {et~al.}(2020)\citenamefont {Barzanjeh}, \citenamefont {Pirandola}, \citenamefont {Vitali},\ and\ \citenamefont {Fink}}]{doi:10.1126/sciadv.abb0451}%
  \BibitemOpen
  \bibfield  {author} {\bibinfo {author} {\bibfnamefont {S.}~\bibnamefont {Barzanjeh}}, \bibinfo {author} {\bibfnamefont {S.}~\bibnamefont {Pirandola}}, \bibinfo {author} {\bibfnamefont {D.}~\bibnamefont {Vitali}},\ and\ \bibinfo {author} {\bibfnamefont {J.~M.}\ \bibnamefont {Fink}},\ }\bibfield  {title} {\bibinfo {title} {Microwave quantum illumination using a digital receiver},\ }\href {https://doi.org/10.1126/sciadv.abb0451} {\bibfield  {journal} {\bibinfo  {journal} {Science Advances}\ }\textbf {\bibinfo {volume} {6}},\ \bibinfo {pages} {eabb0451} (\bibinfo {year} {2020})}\BibitemShut {NoStop}%
\bibitem [{\citenamefont {Chang}\ \emph {et~al.}(2019)\citenamefont {Chang}, \citenamefont {Vadiraj}, \citenamefont {Bourassa}, \citenamefont {Balaji},\ and\ \citenamefont {Wilson}}]{10.1063/1.5085002}%
  \BibitemOpen
  \bibfield  {author} {\bibinfo {author} {\bibfnamefont {C.~W.~S.}\ \bibnamefont {Chang}}, \bibinfo {author} {\bibfnamefont {A.~M.}\ \bibnamefont {Vadiraj}}, \bibinfo {author} {\bibfnamefont {J.}~\bibnamefont {Bourassa}}, \bibinfo {author} {\bibfnamefont {B.}~\bibnamefont {Balaji}},\ and\ \bibinfo {author} {\bibfnamefont {C.~M.}\ \bibnamefont {Wilson}},\ }\bibfield  {title} {\bibinfo {title} {Quantum-enhanced noise radar},\ }\href {https://doi.org/10.1063/1.5085002} {\bibfield  {journal} {\bibinfo  {journal} {Applied Physics Letters}\ }\textbf {\bibinfo {volume} {114}},\ \bibinfo {pages} {112601} (\bibinfo {year} {2019})}\BibitemShut {NoStop}%
\bibitem [{\citenamefont {Luong}\ \emph {et~al.}(2020)\citenamefont {Luong}, \citenamefont {Chang}, \citenamefont {Vadiraj}, \citenamefont {Damini}, \citenamefont {Wilson},\ and\ \citenamefont {Balaji}}]{8890864}%
  \BibitemOpen
  \bibfield  {author} {\bibinfo {author} {\bibfnamefont {D.}~\bibnamefont {Luong}}, \bibinfo {author} {\bibfnamefont {C.~W.~S.}\ \bibnamefont {Chang}}, \bibinfo {author} {\bibfnamefont {A.~M.}\ \bibnamefont {Vadiraj}}, \bibinfo {author} {\bibfnamefont {A.}~\bibnamefont {Damini}}, \bibinfo {author} {\bibfnamefont {C.~M.}\ \bibnamefont {Wilson}},\ and\ \bibinfo {author} {\bibfnamefont {B.}~\bibnamefont {Balaji}},\ }\bibfield  {title} {\bibinfo {title} {Receiver operating characteristics for a prototype quantum two-mode squeezing radar},\ }\href {https://doi.org/10.1109/TAES.2019.2951213} {\bibfield  {journal} {\bibinfo  {journal} {IEEE Transactions on Aerospace and Electronic Systems}\ }\textbf {\bibinfo {volume} {56}},\ \bibinfo {pages} {2041} (\bibinfo {year} {2020})}\BibitemShut {NoStop}%
\bibitem [{\citenamefont {Assouly}\ \emph {et~al.}(2023)\citenamefont {Assouly}, \citenamefont {Dassonneville}, \citenamefont {Peronnin}, \citenamefont {Bienfait},\ and\ \citenamefont {Huard}}]{Assouly2023}%
  \BibitemOpen
  \bibfield  {author} {\bibinfo {author} {\bibfnamefont {R.}~\bibnamefont {Assouly}}, \bibinfo {author} {\bibfnamefont {R.}~\bibnamefont {Dassonneville}}, \bibinfo {author} {\bibfnamefont {T.}~\bibnamefont {Peronnin}}, \bibinfo {author} {\bibfnamefont {A.}~\bibnamefont {Bienfait}},\ and\ \bibinfo {author} {\bibfnamefont {B.}~\bibnamefont {Huard}},\ }\bibfield  {title} {\bibinfo {title} {Quantum advantage in microwave quantum radar},\ }\href {https://doi.org/10.1038/s41567-023-02113-4} {\bibfield  {journal} {\bibinfo  {journal} {Nature Physics}\ }\textbf {\bibinfo {volume} {19}},\ \bibinfo {pages} {1418} (\bibinfo {year} {2023})}\BibitemShut {NoStop}%
\bibitem [{\citenamefont {Zhuang}(2023)}]{Zhuang2023ER}%
  \BibitemOpen
  \bibfield  {author} {\bibinfo {author} {\bibfnamefont {Q.}~\bibnamefont {Zhuang}},\ }\bibfield  {title} {\bibinfo {title} {Quantum advantage on the radar},\ }\href {https://doi.org/10.1038/s41567-023-02111-6} {\bibfield  {journal} {\bibinfo  {journal} {Nature Physics}\ }\textbf {\bibinfo {volume} {19}},\ \bibinfo {pages} {1384} (\bibinfo {year} {2023})}\BibitemShut {NoStop}%
\bibitem [{\citenamefont {{Gallego Torromé}}\ and\ \citenamefont {Barzanjeh}(2024)}]{GALLEGOTORROME2024100497}%
  \BibitemOpen
  \bibfield  {author} {\bibinfo {author} {\bibfnamefont {R.}~\bibnamefont {{Gallego Torromé}}}\ and\ \bibinfo {author} {\bibfnamefont {S.}~\bibnamefont {Barzanjeh}},\ }\bibfield  {title} {\bibinfo {title} {Advances in quantum radar and quantum lidar},\ }\href {https://doi.org/https://doi.org/10.1016/j.pquantelec.2023.100497} {\bibfield  {journal} {\bibinfo  {journal} {Progress in Quantum Electronics}\ }\textbf {\bibinfo {volume} {93}},\ \bibinfo {pages} {100497} (\bibinfo {year} {2024})}\BibitemShut {NoStop}%
\bibitem [{\citenamefont {Karsa}\ \emph {et~al.}(2024)\citenamefont {Karsa}, \citenamefont {Fletcher}, \citenamefont {Spedalieri},\ and\ \citenamefont {Pirandola}}]{Karsa_2024}%
  \BibitemOpen
  \bibfield  {author} {\bibinfo {author} {\bibfnamefont {A.}~\bibnamefont {Karsa}}, \bibinfo {author} {\bibfnamefont {A.}~\bibnamefont {Fletcher}}, \bibinfo {author} {\bibfnamefont {G.}~\bibnamefont {Spedalieri}},\ and\ \bibinfo {author} {\bibfnamefont {S.}~\bibnamefont {Pirandola}},\ }\bibfield  {title} {\bibinfo {title} {Quantum illumination and quantum radar: a brief overview},\ }\href {https://doi.org/10.1088/1361-6633/ad6279} {\bibfield  {journal} {\bibinfo  {journal} {Reports on Progress in Physics}\ }\textbf {\bibinfo {volume} {87}},\ \bibinfo {pages} {094001} (\bibinfo {year} {2024})}\BibitemShut {NoStop}%
\bibitem [{\citenamefont {Zhuang}\ and\ \citenamefont {Shapiro}(2022)}]{PhysRevLett.128.010501}%
  \BibitemOpen
  \bibfield  {author} {\bibinfo {author} {\bibfnamefont {Q.}~\bibnamefont {Zhuang}}\ and\ \bibinfo {author} {\bibfnamefont {J.~H.}\ \bibnamefont {Shapiro}},\ }\bibfield  {title} {\bibinfo {title} {Ultimate accuracy limit of quantum pulse-compression ranging},\ }\href {https://doi.org/10.1103/PhysRevLett.128.010501} {\bibfield  {journal} {\bibinfo  {journal} {Phys. Rev. Lett.}\ }\textbf {\bibinfo {volume} {128}},\ \bibinfo {pages} {010501} (\bibinfo {year} {2022})}\BibitemShut {NoStop}%
\bibitem [{\citenamefont {Reichert}\ \emph {et~al.}(2022)\citenamefont {Reichert}, \citenamefont {Di~Candia}, \citenamefont {Win},\ and\ \citenamefont {Sanz}}]{Reichert2022}%
  \BibitemOpen
  \bibfield  {author} {\bibinfo {author} {\bibfnamefont {M.}~\bibnamefont {Reichert}}, \bibinfo {author} {\bibfnamefont {R.}~\bibnamefont {Di~Candia}}, \bibinfo {author} {\bibfnamefont {M.~Z.}\ \bibnamefont {Win}},\ and\ \bibinfo {author} {\bibfnamefont {M.}~\bibnamefont {Sanz}},\ }\bibfield  {title} {\bibinfo {title} {Quantum-enhanced doppler lidar},\ }\href {https://doi.org/10.1038/s41534-022-00662-9} {\bibfield  {journal} {\bibinfo  {journal} {npj Quantum Information}\ }\textbf {\bibinfo {volume} {8}},\ \bibinfo {pages} {147} (\bibinfo {year} {2022})}\BibitemShut {NoStop}%
\bibitem [{\citenamefont {Cao}\ \emph {et~al.}(2025)\citenamefont {Cao}, \citenamefont {Zhou}, \citenamefont {Zhang}, \citenamefont {Niyato}, \citenamefont {Nie},\ and\ \citenamefont {Han}}]{10872958}%
  \BibitemOpen
  \bibfield  {author} {\bibinfo {author} {\bibfnamefont {J.}~\bibnamefont {Cao}}, \bibinfo {author} {\bibfnamefont {M.}~\bibnamefont {Zhou}}, \bibinfo {author} {\bibfnamefont {R.}~\bibnamefont {Zhang}}, \bibinfo {author} {\bibfnamefont {D.}~\bibnamefont {Niyato}}, \bibinfo {author} {\bibfnamefont {W.}~\bibnamefont {Nie}},\ and\ \bibinfo {author} {\bibfnamefont {Z.}~\bibnamefont {Han}},\ }\bibfield  {title} {\bibinfo {title} {Robust optical quantum imaging framework with entangled photons in oceanic turbulent environments},\ }\href {https://doi.org/10.1109/TCOMM.2025.3538840} {\bibfield  {journal} {\bibinfo  {journal} {IEEE Transactions on Communications}\ }\textbf {\bibinfo {volume} {73}},\ \bibinfo {pages} {6290} (\bibinfo {year} {2025})}\BibitemShut {NoStop}%
\bibitem [{\citenamefont {Zhang}\ \emph {et~al.}(2014)\citenamefont {Zhang}, \citenamefont {Zou}, \citenamefont {Shi}, \citenamefont {Guo},\ and\ \citenamefont {Guo}}]{PhysRevA.90.052308}%
  \BibitemOpen
  \bibfield  {author} {\bibinfo {author} {\bibfnamefont {S.}~\bibnamefont {Zhang}}, \bibinfo {author} {\bibfnamefont {X.}~\bibnamefont {Zou}}, \bibinfo {author} {\bibfnamefont {J.}~\bibnamefont {Shi}}, \bibinfo {author} {\bibfnamefont {J.}~\bibnamefont {Guo}},\ and\ \bibinfo {author} {\bibfnamefont {G.}~\bibnamefont {Guo}},\ }\bibfield  {title} {\bibinfo {title} {Quantum illumination in the presence of photon loss},\ }\href {https://doi.org/10.1103/PhysRevA.90.052308} {\bibfield  {journal} {\bibinfo  {journal} {Phys. Rev. A}\ }\textbf {\bibinfo {volume} {90}},\ \bibinfo {pages} {052308} (\bibinfo {year} {2014})}\BibitemShut {NoStop}%
\bibitem [{\citenamefont {Zhang}\ \emph {et~al.}(2015)\citenamefont {Zhang}, \citenamefont {Mouradian}, \citenamefont {Wong},\ and\ \citenamefont {Shapiro}}]{PhysRevLett.114.110506}%
  \BibitemOpen
  \bibfield  {author} {\bibinfo {author} {\bibfnamefont {Z.}~\bibnamefont {Zhang}}, \bibinfo {author} {\bibfnamefont {S.}~\bibnamefont {Mouradian}}, \bibinfo {author} {\bibfnamefont {F.~N.~C.}\ \bibnamefont {Wong}},\ and\ \bibinfo {author} {\bibfnamefont {J.~H.}\ \bibnamefont {Shapiro}},\ }\bibfield  {title} {\bibinfo {title} {Entanglement-enhanced sensing in a lossy and noisy environment},\ }\href {https://doi.org/10.1103/PhysRevLett.114.110506} {\bibfield  {journal} {\bibinfo  {journal} {Phys. Rev. Lett.}\ }\textbf {\bibinfo {volume} {114}},\ \bibinfo {pages} {110506} (\bibinfo {year} {2015})}\BibitemShut {NoStop}%
\bibitem [{\citenamefont {Jonsson}\ and\ \citenamefont {Candia}(2022)}]{Jonsson_2022}%
  \BibitemOpen
  \bibfield  {author} {\bibinfo {author} {\bibfnamefont {R.}~\bibnamefont {Jonsson}}\ and\ \bibinfo {author} {\bibfnamefont {R.~D.}\ \bibnamefont {Candia}},\ }\bibfield  {title} {\bibinfo {title} {Gaussian quantum estimation of the loss parameter in a thermal environment},\ }\href {https://doi.org/10.1088/1751-8121/ac83fa} {\bibfield  {journal} {\bibinfo  {journal} {Journal of Physics A: Mathematical and Theoretical}\ }\textbf {\bibinfo {volume} {55}},\ \bibinfo {pages} {385301} (\bibinfo {year} {2022})}\BibitemShut {NoStop}%
\bibitem [{\citenamefont {Park}\ \emph {et~al.}(2023)\citenamefont {Park}, \citenamefont {Noh},\ and\ \citenamefont {Lee}}]{Park2023}%
  \BibitemOpen
  \bibfield  {author} {\bibinfo {author} {\bibfnamefont {S.-i.}\ \bibnamefont {Park}}, \bibinfo {author} {\bibfnamefont {C.}~\bibnamefont {Noh}},\ and\ \bibinfo {author} {\bibfnamefont {C.}~\bibnamefont {Lee}},\ }\bibfield  {title} {\bibinfo {title} {Quantum loss sensing with two-mode squeezed vacuum state under noisy and lossy environment},\ }\href {https://doi.org/10.1038/s41598-023-32770-7} {\bibfield  {journal} {\bibinfo  {journal} {Scientific Reports}\ }\textbf {\bibinfo {volume} {13}},\ \bibinfo {pages} {5936} (\bibinfo {year} {2023})}\BibitemShut {NoStop}%
\bibitem [{\citenamefont {Kim}\ \emph {et~al.}(2023)\citenamefont {Kim}, \citenamefont {Jo}, \citenamefont {Kim}, \citenamefont {Jeong}, \citenamefont {Kim}, \citenamefont {Park}, \citenamefont {Kim},\ and\ \citenamefont {Lee}}]{PhysRevResearch.5.033010}%
  \BibitemOpen
  \bibfield  {author} {\bibinfo {author} {\bibfnamefont {D.~H.}\ \bibnamefont {Kim}}, \bibinfo {author} {\bibfnamefont {Y.}~\bibnamefont {Jo}}, \bibinfo {author} {\bibfnamefont {D.~Y.}\ \bibnamefont {Kim}}, \bibinfo {author} {\bibfnamefont {T.}~\bibnamefont {Jeong}}, \bibinfo {author} {\bibfnamefont {J.}~\bibnamefont {Kim}}, \bibinfo {author} {\bibfnamefont {N.~H.}\ \bibnamefont {Park}}, \bibinfo {author} {\bibfnamefont {Z.}~\bibnamefont {Kim}},\ and\ \bibinfo {author} {\bibfnamefont {S.-Y.}\ \bibnamefont {Lee}},\ }\bibfield  {title} {\bibinfo {title} {Gaussian quantum illumination via monotone metrics},\ }\href {https://doi.org/10.1103/PhysRevResearch.5.033010} {\bibfield  {journal} {\bibinfo  {journal} {Phys. Rev. Res.}\ }\textbf {\bibinfo {volume} {5}},\ \bibinfo {pages} {033010} (\bibinfo {year} {2023})}\BibitemShut {NoStop}%
\bibitem [{\citenamefont {Zhang}\ \emph {et~al.}(2025{\natexlab{a}})\citenamefont {Zhang}, \citenamefont {Li}, \citenamefont {Liu}, \citenamefont {Li}, \citenamefont {Sun},\ and\ \citenamefont {Yu}}]{PhysRevA.111.022627}%
  \BibitemOpen
  \bibfield  {author} {\bibinfo {author} {\bibfnamefont {S.}~\bibnamefont {Zhang}}, \bibinfo {author} {\bibfnamefont {L.}~\bibnamefont {Li}}, \bibinfo {author} {\bibfnamefont {M.}~\bibnamefont {Liu}}, \bibinfo {author} {\bibfnamefont {J.}~\bibnamefont {Li}}, \bibinfo {author} {\bibfnamefont {W.}~\bibnamefont {Sun}},\ and\ \bibinfo {author} {\bibfnamefont {Q.}~\bibnamefont {Yu}},\ }\bibfield  {title} {\bibinfo {title} {Enhancing the performance of quantum radar with zero-photon subtraction},\ }\href {https://doi.org/10.1103/PhysRevA.111.022627} {\bibfield  {journal} {\bibinfo  {journal} {Phys. Rev. A}\ }\textbf {\bibinfo {volume} {111}},\ \bibinfo {pages} {022627} (\bibinfo {year} {2025}{\natexlab{a}})}\BibitemShut {NoStop}%
\bibitem [{\citenamefont {Monken}\ \emph {et~al.}(1998)\citenamefont {Monken}, \citenamefont {Ribeiro},\ and\ \citenamefont {P\'adua}}]{PhysRevA.57.3123}%
  \BibitemOpen
  \bibfield  {author} {\bibinfo {author} {\bibfnamefont {C.~H.}\ \bibnamefont {Monken}}, \bibinfo {author} {\bibfnamefont {P.~H.~S.}\ \bibnamefont {Ribeiro}},\ and\ \bibinfo {author} {\bibfnamefont {S.}~\bibnamefont {P\'adua}},\ }\bibfield  {title} {\bibinfo {title} {Transfer of angular spectrum and image formation in spontaneous parametric down-conversion},\ }\href {https://doi.org/10.1103/PhysRevA.57.3123} {\bibfield  {journal} {\bibinfo  {journal} {Phys. Rev. A}\ }\textbf {\bibinfo {volume} {57}},\ \bibinfo {pages} {3123} (\bibinfo {year} {1998})}\BibitemShut {NoStop}%
\bibitem [{\citenamefont {Valencia}\ \emph {et~al.}(2007)\citenamefont {Valencia}, \citenamefont {Cer\'e}, \citenamefont {Shi}, \citenamefont {Molina-Terriza},\ and\ \citenamefont {Torres}}]{PhysRevLett.99.243601}%
  \BibitemOpen
  \bibfield  {author} {\bibinfo {author} {\bibfnamefont {A.}~\bibnamefont {Valencia}}, \bibinfo {author} {\bibfnamefont {A.}~\bibnamefont {Cer\'e}}, \bibinfo {author} {\bibfnamefont {X.}~\bibnamefont {Shi}}, \bibinfo {author} {\bibfnamefont {G.}~\bibnamefont {Molina-Terriza}},\ and\ \bibinfo {author} {\bibfnamefont {J.~P.}\ \bibnamefont {Torres}},\ }\bibfield  {title} {\bibinfo {title} {Shaping the waveform of entangled photons},\ }\href {https://doi.org/10.1103/PhysRevLett.99.243601} {\bibfield  {journal} {\bibinfo  {journal} {Phys. Rev. Lett.}\ }\textbf {\bibinfo {volume} {99}},\ \bibinfo {pages} {243601} (\bibinfo {year} {2007})}\BibitemShut {NoStop}%
\bibitem [{\citenamefont {Boucher}\ \emph {et~al.}(2015)\citenamefont {Boucher}, \citenamefont {Douce}, \citenamefont {Bresteau}, \citenamefont {Walborn}, \citenamefont {Keller}, \citenamefont {Coudreau}, \citenamefont {Ducci},\ and\ \citenamefont {Milman}}]{PhysRevA.92.023804}%
  \BibitemOpen
  \bibfield  {author} {\bibinfo {author} {\bibfnamefont {G.}~\bibnamefont {Boucher}}, \bibinfo {author} {\bibfnamefont {T.}~\bibnamefont {Douce}}, \bibinfo {author} {\bibfnamefont {D.}~\bibnamefont {Bresteau}}, \bibinfo {author} {\bibfnamefont {S.~P.}\ \bibnamefont {Walborn}}, \bibinfo {author} {\bibfnamefont {A.}~\bibnamefont {Keller}}, \bibinfo {author} {\bibfnamefont {T.}~\bibnamefont {Coudreau}}, \bibinfo {author} {\bibfnamefont {S.}~\bibnamefont {Ducci}},\ and\ \bibinfo {author} {\bibfnamefont {P.}~\bibnamefont {Milman}},\ }\bibfield  {title} {\bibinfo {title} {Toolbox for continuous-variable entanglement production and measurement using spontaneous parametric down-conversion},\ }\href {https://doi.org/10.1103/PhysRevA.92.023804} {\bibfield  {journal} {\bibinfo  {journal} {Phys. Rev. A}\ }\textbf {\bibinfo {volume} {92}},\ \bibinfo {pages} {023804} (\bibinfo {year} {2015})}\BibitemShut {NoStop}%
\bibitem [{\citenamefont {Sofer}\ \emph {et~al.}(2019)\citenamefont {Sofer}, \citenamefont {Strizhevsky}, \citenamefont {Schori}, \citenamefont {Tamasaku},\ and\ \citenamefont {Shwartz}}]{PhysRevX.9.031033}%
  \BibitemOpen
  \bibfield  {author} {\bibinfo {author} {\bibfnamefont {S.}~\bibnamefont {Sofer}}, \bibinfo {author} {\bibfnamefont {E.}~\bibnamefont {Strizhevsky}}, \bibinfo {author} {\bibfnamefont {A.}~\bibnamefont {Schori}}, \bibinfo {author} {\bibfnamefont {K.}~\bibnamefont {Tamasaku}},\ and\ \bibinfo {author} {\bibfnamefont {S.}~\bibnamefont {Shwartz}},\ }\bibfield  {title} {\bibinfo {title} {Quantum enhanced x-ray detection},\ }\href {https://doi.org/10.1103/PhysRevX.9.031033} {\bibfield  {journal} {\bibinfo  {journal} {Phys. Rev. X}\ }\textbf {\bibinfo {volume} {9}},\ \bibinfo {pages} {031033} (\bibinfo {year} {2019})}\BibitemShut {NoStop}%
\bibitem [{\citenamefont {Pirandola}\ and\ \citenamefont {Lloyd}(2008)}]{PhysRevA.78.012331}%
  \BibitemOpen
  \bibfield  {author} {\bibinfo {author} {\bibfnamefont {S.}~\bibnamefont {Pirandola}}\ and\ \bibinfo {author} {\bibfnamefont {S.}~\bibnamefont {Lloyd}},\ }\bibfield  {title} {\bibinfo {title} {Computable bounds for the discrimination of gaussian states},\ }\href {https://doi.org/10.1103/PhysRevA.78.012331} {\bibfield  {journal} {\bibinfo  {journal} {Phys. Rev. A}\ }\textbf {\bibinfo {volume} {78}},\ \bibinfo {pages} {012331} (\bibinfo {year} {2008})}\BibitemShut {NoStop}%
\bibitem [{\citenamefont {Weedbrook}\ \emph {et~al.}(2012)\citenamefont {Weedbrook}, \citenamefont {Pirandola}, \citenamefont {Garc\'{\i}a-Patr\'on}, \citenamefont {Cerf}, \citenamefont {Ralph}, \citenamefont {Shapiro},\ and\ \citenamefont {Lloyd}}]{RevModPhys.84.621}%
  \BibitemOpen
  \bibfield  {author} {\bibinfo {author} {\bibfnamefont {C.}~\bibnamefont {Weedbrook}}, \bibinfo {author} {\bibfnamefont {S.}~\bibnamefont {Pirandola}}, \bibinfo {author} {\bibfnamefont {R.}~\bibnamefont {Garc\'{\i}a-Patr\'on}}, \bibinfo {author} {\bibfnamefont {N.~J.}\ \bibnamefont {Cerf}}, \bibinfo {author} {\bibfnamefont {T.~C.}\ \bibnamefont {Ralph}}, \bibinfo {author} {\bibfnamefont {J.~H.}\ \bibnamefont {Shapiro}},\ and\ \bibinfo {author} {\bibfnamefont {S.}~\bibnamefont {Lloyd}},\ }\bibfield  {title} {\bibinfo {title} {Gaussian quantum information},\ }\href {https://doi.org/10.1103/RevModPhys.84.621} {\bibfield  {journal} {\bibinfo  {journal} {Rev. Mod. Phys.}\ }\textbf {\bibinfo {volume} {84}},\ \bibinfo {pages} {621} (\bibinfo {year} {2012})}\BibitemShut {NoStop}%
\bibitem [{\citenamefont {He}\ \emph {et~al.}(2024)\citenamefont {He}, \citenamefont {Feng}, \citenamefont {Wu}, \citenamefont {Song},\ and\ \citenamefont {Wei}}]{PhysRevA.110.052403}%
  \BibitemOpen
  \bibfield  {author} {\bibinfo {author} {\bibfnamefont {D.}~\bibnamefont {He}}, \bibinfo {author} {\bibfnamefont {X.~N.}\ \bibnamefont {Feng}}, \bibinfo {author} {\bibfnamefont {Y.}~\bibnamefont {Wu}}, \bibinfo {author} {\bibfnamefont {H.}~\bibnamefont {Song}},\ and\ \bibinfo {author} {\bibfnamefont {L.~F.}\ \bibnamefont {Wei}},\ }\bibfield  {title} {\bibinfo {title} {Enhancing the achievable sensitivity of coherent-state illumination by using photon-number-resolvable detection},\ }\href {https://doi.org/10.1103/PhysRevA.110.052403} {\bibfield  {journal} {\bibinfo  {journal} {Phys. Rev. A}\ }\textbf {\bibinfo {volume} {110}},\ \bibinfo {pages} {052403} (\bibinfo {year} {2024})}\BibitemShut {NoStop}%
\bibitem [{\citenamefont {Tham}\ \emph {et~al.}(2024)\citenamefont {Tham}, \citenamefont {Nair},\ and\ \citenamefont {Gu}}]{PhysRevLett.133.110801}%
  \BibitemOpen
  \bibfield  {author} {\bibinfo {author} {\bibfnamefont {G.~Y.}\ \bibnamefont {Tham}}, \bibinfo {author} {\bibfnamefont {R.}~\bibnamefont {Nair}},\ and\ \bibinfo {author} {\bibfnamefont {M.}~\bibnamefont {Gu}},\ }\bibfield  {title} {\bibinfo {title} {Quantum limits of covert target detection},\ }\href {https://doi.org/10.1103/PhysRevLett.133.110801} {\bibfield  {journal} {\bibinfo  {journal} {Phys. Rev. Lett.}\ }\textbf {\bibinfo {volume} {133}},\ \bibinfo {pages} {110801} (\bibinfo {year} {2024})}\BibitemShut {NoStop}%
\bibitem [{\citenamefont {Calsamiglia}\ \emph {et~al.}(2008)\citenamefont {Calsamiglia}, \citenamefont {Mu\~noz Tapia}, \citenamefont {Masanes}, \citenamefont {Acin},\ and\ \citenamefont {Bagan}}]{PhysRevA.77.032311}%
  \BibitemOpen
  \bibfield  {author} {\bibinfo {author} {\bibfnamefont {J.}~\bibnamefont {Calsamiglia}}, \bibinfo {author} {\bibfnamefont {R.}~\bibnamefont {Mu\~noz Tapia}}, \bibinfo {author} {\bibfnamefont {L.}~\bibnamefont {Masanes}}, \bibinfo {author} {\bibfnamefont {A.}~\bibnamefont {Acin}},\ and\ \bibinfo {author} {\bibfnamefont {E.}~\bibnamefont {Bagan}},\ }\bibfield  {title} {\bibinfo {title} {Quantum chernoff bound as a measure of distinguishability between density matrices: Application to qubit and gaussian states},\ }\href {https://doi.org/10.1103/PhysRevA.77.032311} {\bibfield  {journal} {\bibinfo  {journal} {Phys. Rev. A}\ }\textbf {\bibinfo {volume} {77}},\ \bibinfo {pages} {032311} (\bibinfo {year} {2008})}\BibitemShut {NoStop}%
\bibitem [{\citenamefont {Banchi}\ \emph {et~al.}(2015)\citenamefont {Banchi}, \citenamefont {Braunstein},\ and\ \citenamefont {Pirandola}}]{PhysRevLett.115.260501}%
  \BibitemOpen
  \bibfield  {author} {\bibinfo {author} {\bibfnamefont {L.}~\bibnamefont {Banchi}}, \bibinfo {author} {\bibfnamefont {S.~L.}\ \bibnamefont {Braunstein}},\ and\ \bibinfo {author} {\bibfnamefont {S.}~\bibnamefont {Pirandola}},\ }\bibfield  {title} {\bibinfo {title} {Quantum fidelity for arbitrary gaussian states},\ }\href {https://doi.org/10.1103/PhysRevLett.115.260501} {\bibfield  {journal} {\bibinfo  {journal} {Phys. Rev. Lett.}\ }\textbf {\bibinfo {volume} {115}},\ \bibinfo {pages} {260501} (\bibinfo {year} {2015})}\BibitemShut {NoStop}%
\bibitem [{\citenamefont {Marian}\ and\ \citenamefont {Marian}(2012)}]{PhysRevA.86.022340}%
  \BibitemOpen
  \bibfield  {author} {\bibinfo {author} {\bibfnamefont {P.}~\bibnamefont {Marian}}\ and\ \bibinfo {author} {\bibfnamefont {T.~A.}\ \bibnamefont {Marian}},\ }\bibfield  {title} {\bibinfo {title} {Uhlmann fidelity between two-mode gaussian states},\ }\href {https://doi.org/10.1103/PhysRevA.86.022340} {\bibfield  {journal} {\bibinfo  {journal} {Phys. Rev. A}\ }\textbf {\bibinfo {volume} {86}},\ \bibinfo {pages} {022340} (\bibinfo {year} {2012})}\BibitemShut {NoStop}%
\bibitem [{\citenamefont {Marian}\ and\ \citenamefont {Marian}(2016)}]{PhysRevA.93.052330}%
  \BibitemOpen
  \bibfield  {author} {\bibinfo {author} {\bibfnamefont {P.}~\bibnamefont {Marian}}\ and\ \bibinfo {author} {\bibfnamefont {T.~A.}\ \bibnamefont {Marian}},\ }\bibfield  {title} {\bibinfo {title} {Quantum fisher information on two manifolds of two-mode gaussian states},\ }\href {https://doi.org/10.1103/PhysRevA.93.052330} {\bibfield  {journal} {\bibinfo  {journal} {Phys. Rev. A}\ }\textbf {\bibinfo {volume} {93}},\ \bibinfo {pages} {052330} (\bibinfo {year} {2016})}\BibitemShut {NoStop}%
\bibitem [{\citenamefont {Leggett}\ \emph {et~al.}(1987)\citenamefont {Leggett}, \citenamefont {Chakravarty}, \citenamefont {Dorsey}, \citenamefont {Fisher}, \citenamefont {Garg},\ and\ \citenamefont {Zwerger}}]{RevModPhys.59.1}%
  \BibitemOpen
  \bibfield  {author} {\bibinfo {author} {\bibfnamefont {A.~J.}\ \bibnamefont {Leggett}}, \bibinfo {author} {\bibfnamefont {S.}~\bibnamefont {Chakravarty}}, \bibinfo {author} {\bibfnamefont {A.~T.}\ \bibnamefont {Dorsey}}, \bibinfo {author} {\bibfnamefont {M.~P.~A.}\ \bibnamefont {Fisher}}, \bibinfo {author} {\bibfnamefont {A.}~\bibnamefont {Garg}},\ and\ \bibinfo {author} {\bibfnamefont {W.}~\bibnamefont {Zwerger}},\ }\bibfield  {title} {\bibinfo {title} {Dynamics of the dissipative two-state system},\ }\href {https://doi.org/10.1103/RevModPhys.59.1} {\bibfield  {journal} {\bibinfo  {journal} {Rev. Mod. Phys.}\ }\textbf {\bibinfo {volume} {59}},\ \bibinfo {pages} {1} (\bibinfo {year} {1987})}\BibitemShut {NoStop}%
\bibitem [{\citenamefont {An}\ and\ \citenamefont {Zhang}(2007)}]{PhysRevA.76.042127}%
  \BibitemOpen
  \bibfield  {author} {\bibinfo {author} {\bibfnamefont {J.-H.}\ \bibnamefont {An}}\ and\ \bibinfo {author} {\bibfnamefont {W.-M.}\ \bibnamefont {Zhang}},\ }\bibfield  {title} {\bibinfo {title} {Non-markovian entanglement dynamics of noisy continuous-variable quantum channels},\ }\href {https://doi.org/10.1103/PhysRevA.76.042127} {\bibfield  {journal} {\bibinfo  {journal} {Phys. Rev. A}\ }\textbf {\bibinfo {volume} {76}},\ \bibinfo {pages} {042127} (\bibinfo {year} {2007})}\BibitemShut {NoStop}%
\bibitem [{\citenamefont {An}\ \emph {et~al.}(2008)\citenamefont {An}, \citenamefont {Yeo}, \citenamefont {Zhang},\ and\ \citenamefont {Oh}}]{An_2009}%
  \BibitemOpen
  \bibfield  {author} {\bibinfo {author} {\bibfnamefont {J.-H.}\ \bibnamefont {An}}, \bibinfo {author} {\bibfnamefont {Y.}~\bibnamefont {Yeo}}, \bibinfo {author} {\bibfnamefont {W.-M.}\ \bibnamefont {Zhang}},\ and\ \bibinfo {author} {\bibfnamefont {C.~H.}\ \bibnamefont {Oh}},\ }\bibfield  {title} {\bibinfo {title} {Entanglement oscillation and survival induced by non-markovian decoherence dynamics of the entangled squeezed state},\ }\href {https://doi.org/10.1088/1751-8113/42/1/015302} {\bibfield  {journal} {\bibinfo  {journal} {Journal of Physics A: Mathematical and Theoretical}\ }\textbf {\bibinfo {volume} {42}},\ \bibinfo {pages} {015302} (\bibinfo {year} {2008})}\BibitemShut {NoStop}%
\bibitem [{\citenamefont {L\"u}\ \emph {et~al.}(2013)\citenamefont {L\"u}, \citenamefont {An}, \citenamefont {Chen}, \citenamefont {Luo},\ and\ \citenamefont {Oh}}]{PhysRevA.88.012129}%
  \BibitemOpen
  \bibfield  {author} {\bibinfo {author} {\bibfnamefont {Y.-Q.}\ \bibnamefont {L\"u}}, \bibinfo {author} {\bibfnamefont {J.-H.}\ \bibnamefont {An}}, \bibinfo {author} {\bibfnamefont {X.-M.}\ \bibnamefont {Chen}}, \bibinfo {author} {\bibfnamefont {H.-G.}\ \bibnamefont {Luo}},\ and\ \bibinfo {author} {\bibfnamefont {C.~H.}\ \bibnamefont {Oh}},\ }\bibfield  {title} {\bibinfo {title} {Frozen gaussian quantum discord in photonic crystal cavity array system},\ }\href {https://doi.org/10.1103/PhysRevA.88.012129} {\bibfield  {journal} {\bibinfo  {journal} {Phys. Rev. A}\ }\textbf {\bibinfo {volume} {88}},\ \bibinfo {pages} {012129} (\bibinfo {year} {2013})}\BibitemShut {NoStop}%
\bibitem [{Sup()}]{SupplementalMaterial}%
  \BibitemOpen
  \href@noop {} {}\bibinfo {note} {See the Supplementary Materials for more details on the derivation of the exact master equation, the covariance matrix and the full expression of the quantum fidelity.}\BibitemShut {Stop}%
\bibitem [{\citenamefont {Yang}\ \emph {et~al.}(2014)\citenamefont {Yang}, \citenamefont {An}, \citenamefont {Luo}, \citenamefont {Li},\ and\ \citenamefont {Oh}}]{PhysRevE.90.022122}%
  \BibitemOpen
  \bibfield  {author} {\bibinfo {author} {\bibfnamefont {C.-J.}\ \bibnamefont {Yang}}, \bibinfo {author} {\bibfnamefont {J.-H.}\ \bibnamefont {An}}, \bibinfo {author} {\bibfnamefont {H.-G.}\ \bibnamefont {Luo}}, \bibinfo {author} {\bibfnamefont {Y.}~\bibnamefont {Li}},\ and\ \bibinfo {author} {\bibfnamefont {C.~H.}\ \bibnamefont {Oh}},\ }\bibfield  {title} {\bibinfo {title} {Canonical versus noncanonical equilibration dynamics of open quantum systems},\ }\href {https://doi.org/10.1103/PhysRevE.90.022122} {\bibfield  {journal} {\bibinfo  {journal} {Phys. Rev. E}\ }\textbf {\bibinfo {volume} {90}},\ \bibinfo {pages} {022122} (\bibinfo {year} {2014})}\BibitemShut {NoStop}%
\bibitem [{\citenamefont {Zhang}\ \emph {et~al.}(2025{\natexlab{b}})\citenamefont {Zhang}, \citenamefont {Zheng}, \citenamefont {Wang},\ and\ \citenamefont {Zhang}}]{Zhang_2025}%
  \BibitemOpen
  \bibfield  {author} {\bibinfo {author} {\bibfnamefont {J.}~\bibnamefont {Zhang}}, \bibinfo {author} {\bibfnamefont {K.}~\bibnamefont {Zheng}}, \bibinfo {author} {\bibfnamefont {B.}~\bibnamefont {Wang}},\ and\ \bibinfo {author} {\bibfnamefont {L.}~\bibnamefont {Zhang}},\ }\bibfield  {title} {\bibinfo {title} {Performance advantage of quantum illumination using symmetric and asymmetric hypothesis testings in lossy environments},\ }\href {https://doi.org/10.1088/1751-8121/addb94} {\bibfield  {journal} {\bibinfo  {journal} {Journal of Physics A: Mathematical and Theoretical}\ }\textbf {\bibinfo {volume} {58}},\ \bibinfo {pages} {225301} (\bibinfo {year} {2025}{\natexlab{b}})}\BibitemShut {NoStop}%
\bibitem [{\citenamefont {Wu}\ \emph {et~al.}(2021)\citenamefont {Wu}, \citenamefont {Bai},\ and\ \citenamefont {An}}]{PhysRevA.103.L010601}%
  \BibitemOpen
  \bibfield  {author} {\bibinfo {author} {\bibfnamefont {W.}~\bibnamefont {Wu}}, \bibinfo {author} {\bibfnamefont {S.-Y.}\ \bibnamefont {Bai}},\ and\ \bibinfo {author} {\bibfnamefont {J.-H.}\ \bibnamefont {An}},\ }\bibfield  {title} {\bibinfo {title} {Non-markovian sensing of a quantum reservoir},\ }\href {https://doi.org/10.1103/PhysRevA.103.L010601} {\bibfield  {journal} {\bibinfo  {journal} {Phys. Rev. A}\ }\textbf {\bibinfo {volume} {103}},\ \bibinfo {pages} {L010601} (\bibinfo {year} {2021})}\BibitemShut {NoStop}%
\bibitem [{\citenamefont {Myatt}\ \emph {et~al.}(2000)\citenamefont {Myatt}, \citenamefont {King}, \citenamefont {Turchette}, \citenamefont {Sackett}, \citenamefont {Kielpinski}, \citenamefont {Itano}, \citenamefont {Monroe},\ and\ \citenamefont {Wineland}}]{ER1}%
  \BibitemOpen
  \bibfield  {author} {\bibinfo {author} {\bibfnamefont {C.~J.}\ \bibnamefont {Myatt}}, \bibinfo {author} {\bibfnamefont {B.~E.}\ \bibnamefont {King}}, \bibinfo {author} {\bibfnamefont {Q.~A.}\ \bibnamefont {Turchette}}, \bibinfo {author} {\bibfnamefont {C.~A.}\ \bibnamefont {Sackett}}, \bibinfo {author} {\bibfnamefont {D.}~\bibnamefont {Kielpinski}}, \bibinfo {author} {\bibfnamefont {W.~M.}\ \bibnamefont {Itano}}, \bibinfo {author} {\bibfnamefont {C.}~\bibnamefont {Monroe}},\ and\ \bibinfo {author} {\bibfnamefont {D.~J.}\ \bibnamefont {Wineland}},\ }\bibfield  {title} {\bibinfo {title} {Decoherence of quantum superpositions through coupling to engineered reservoirs},\ }\href {https://doi.org/10.1038/35002001} {\bibfield  {journal} {\bibinfo  {journal} {Nature}\ }\textbf {\bibinfo {volume} {403}},\ \bibinfo {pages} {269} (\bibinfo {year} {2000})}\BibitemShut {NoStop}%
\bibitem [{\citenamefont {Kienzler}\ \emph {et~al.}(2015)\citenamefont {Kienzler}, \citenamefont {Lo}, \citenamefont {Keitch}, \citenamefont {de~Clercq}, \citenamefont {Leupold}, \citenamefont {Lindenfelser}, \citenamefont {Marinelli}, \citenamefont {Negnevitsky},\ and\ \citenamefont {Home}}]{Kienzler53}%
  \BibitemOpen
  \bibfield  {author} {\bibinfo {author} {\bibfnamefont {D.}~\bibnamefont {Kienzler}}, \bibinfo {author} {\bibfnamefont {H.-Y.}\ \bibnamefont {Lo}}, \bibinfo {author} {\bibfnamefont {B.}~\bibnamefont {Keitch}}, \bibinfo {author} {\bibfnamefont {L.}~\bibnamefont {de~Clercq}}, \bibinfo {author} {\bibfnamefont {F.}~\bibnamefont {Leupold}}, \bibinfo {author} {\bibfnamefont {F.}~\bibnamefont {Lindenfelser}}, \bibinfo {author} {\bibfnamefont {M.}~\bibnamefont {Marinelli}}, \bibinfo {author} {\bibfnamefont {V.}~\bibnamefont {Negnevitsky}},\ and\ \bibinfo {author} {\bibfnamefont {J.~P.}\ \bibnamefont {Home}},\ }\bibfield  {title} {\bibinfo {title} {Quantum harmonic oscillator state synthesis by reservoir engineering},\ }\href {https://doi.org/10.1126/science.1261033} {\bibfield  {journal} {\bibinfo  {journal} {Science}\ }\textbf {\bibinfo {volume} {347}},\ \bibinfo {pages} {53} (\bibinfo {year} {2015})}\BibitemShut {NoStop}%
\bibitem [{\citenamefont {Porras}\ \emph {et~al.}(2008)\citenamefont {Porras}, \citenamefont {Marquardt}, \citenamefont {von Delft},\ and\ \citenamefont {Cirac}}]{PhysRevA.78.010101}%
  \BibitemOpen
  \bibfield  {author} {\bibinfo {author} {\bibfnamefont {D.}~\bibnamefont {Porras}}, \bibinfo {author} {\bibfnamefont {F.}~\bibnamefont {Marquardt}}, \bibinfo {author} {\bibfnamefont {J.}~\bibnamefont {von Delft}},\ and\ \bibinfo {author} {\bibfnamefont {J.~I.}\ \bibnamefont {Cirac}},\ }\bibfield  {title} {\bibinfo {title} {Mesoscopic spin-boson models of trapped ions},\ }\href {https://doi.org/10.1103/PhysRevA.78.010101} {\bibfield  {journal} {\bibinfo  {journal} {Phys. Rev. A}\ }\textbf {\bibinfo {volume} {78}},\ \bibinfo {pages} {010101(R)} (\bibinfo {year} {2008})}\BibitemShut {NoStop}%
\bibitem [{\citenamefont {Kitzman}\ \emph {et~al.}(2023)\citenamefont {Kitzman}, \citenamefont {Lane}, \citenamefont {Undershute}, \citenamefont {Harrington}, \citenamefont {Beysengulov}, \citenamefont {Mikolas}, \citenamefont {Murch},\ and\ \citenamefont {Pollanen}}]{Kitzman2023}%
  \BibitemOpen
  \bibfield  {author} {\bibinfo {author} {\bibfnamefont {J.~M.}\ \bibnamefont {Kitzman}}, \bibinfo {author} {\bibfnamefont {J.~R.}\ \bibnamefont {Lane}}, \bibinfo {author} {\bibfnamefont {C.}~\bibnamefont {Undershute}}, \bibinfo {author} {\bibfnamefont {P.~M.}\ \bibnamefont {Harrington}}, \bibinfo {author} {\bibfnamefont {N.~R.}\ \bibnamefont {Beysengulov}}, \bibinfo {author} {\bibfnamefont {C.~A.}\ \bibnamefont {Mikolas}}, \bibinfo {author} {\bibfnamefont {K.~W.}\ \bibnamefont {Murch}},\ and\ \bibinfo {author} {\bibfnamefont {J.}~\bibnamefont {Pollanen}},\ }\bibfield  {title} {\bibinfo {title} {Phononic bath engineering of a superconducting qubit},\ }\href {https://doi.org/10.1038/s41467-023-39682-0} {\bibfield  {journal} {\bibinfo  {journal} {Nature Communications}\ }\textbf {\bibinfo {volume} {14}},\ \bibinfo {pages} {3910} (\bibinfo {year} {2023})}\BibitemShut {NoStop}%
\bibitem [{\citenamefont {Liu}\ and\ \citenamefont {Houck}(2017)}]{Liu2017}%
  \BibitemOpen
  \bibfield  {author} {\bibinfo {author} {\bibfnamefont {Y.}~\bibnamefont {Liu}}\ and\ \bibinfo {author} {\bibfnamefont {A.~A.}\ \bibnamefont {Houck}},\ }\bibfield  {title} {\bibinfo {title} {Quantum electrodynamics near a photonic bandgap},\ }\href {https://doi.org/10.1038/nphys3834} {\bibfield  {journal} {\bibinfo  {journal} {Nature Physics}\ }\textbf {\bibinfo {volume} {13}},\ \bibinfo {pages} {48} (\bibinfo {year} {2017})}\BibitemShut {NoStop}%
\bibitem [{\citenamefont {Krinner}\ \emph {et~al.}(2018)\citenamefont {Krinner}, \citenamefont {Stewart}, \citenamefont {Pazmi{\~{n}}o}, \citenamefont {Kwon},\ and\ \citenamefont {Schneble}}]{Krinner2018}%
  \BibitemOpen
  \bibfield  {author} {\bibinfo {author} {\bibfnamefont {L.}~\bibnamefont {Krinner}}, \bibinfo {author} {\bibfnamefont {M.}~\bibnamefont {Stewart}}, \bibinfo {author} {\bibfnamefont {A.}~\bibnamefont {Pazmi{\~{n}}o}}, \bibinfo {author} {\bibfnamefont {J.}~\bibnamefont {Kwon}},\ and\ \bibinfo {author} {\bibfnamefont {D.}~\bibnamefont {Schneble}},\ }\bibfield  {title} {\bibinfo {title} {Spontaneous emission of matter waves from a tunable open quantum system},\ }\href {https://doi.org/10.1038/s41586-018-0348-z} {\bibfield  {journal} {\bibinfo  {journal} {Nature}\ }\textbf {\bibinfo {volume} {559}},\ \bibinfo {pages} {589} (\bibinfo {year} {2018})}\BibitemShut {NoStop}%
\bibitem [{\citenamefont {Kwon}\ \emph {et~al.}(2022)\citenamefont {Kwon}, \citenamefont {Kim}, \citenamefont {Lanuza},\ and\ \citenamefont {Schneble}}]{RN11}%
  \BibitemOpen
  \bibfield  {author} {\bibinfo {author} {\bibfnamefont {J.}~\bibnamefont {Kwon}}, \bibinfo {author} {\bibfnamefont {Y.}~\bibnamefont {Kim}}, \bibinfo {author} {\bibfnamefont {A.}~\bibnamefont {Lanuza}},\ and\ \bibinfo {author} {\bibfnamefont {D.}~\bibnamefont {Schneble}},\ }\bibfield  {title} {\bibinfo {title} {Formation of matter-wave polaritons in an optical lattice},\ }\href {https://doi.org/10.1038/s41567-022-01565-4} {\bibfield  {journal} {\bibinfo  {journal} {Nature Physics}\ }\textbf {\bibinfo {volume} {18}},\ \bibinfo {pages} {657} (\bibinfo {year} {2022})}\BibitemShut {NoStop}%
\end{thebibliography}%
\end{document}